\documentclass[12pt]{article}
\usepackage{graphicx}
\usepackage[a4paper, total={6in, 9in}]{geometry}
\usepackage{indentfirst,csquotes}
\usepackage{enumitem}
\usepackage{setspace}
\usepackage{algorithm}
\usepackage{algorithmic}
\usepackage{bm}
\usepackage{bbm}
\usepackage{natbib}

\usepackage{amssymb,amsthm,amsmath}
\usepackage{xcolor,hyperref,titlesec,fancyhdr,etoolbox}
\usepackage{subfigure}
\usepackage{longtable}

\usepackage{cleveref}
\usepackage[]{natbib} % Square brackets, citing references with numbers, citations sorted by appearance in the text and compressed
\usepackage{authblk}

\begin{document}
\doublespacing
\title{Ordinal Mixed-Effects Random Forest}%\CM{QUASI QUASI TOGLIEREI an innovative statistical method to perform learning analytics}} 
\date{}

\author[1]{Giulia Bergonzoli  \thanks{Email: giulia.bergonzoli@gmail.com}}
\author[1,2]{Lidia Rossi \thanks{Email: lidia.rossi@polimi.it}}
\author[1]{Chiara Masci \thanks{Email: chiara.masci@polimi.it}}
\affil[1]{\small MOX Laboratory, Department of Mathematics, Politecnico di Milano, Milano, Italy}
\affil[2]{\small Department of Management, Economics and Industrial Engineering, Politecnico di Milano, Milano, Italy}

\maketitle
\vspace{-0.8cm}
\begin{abstract}
%In this work
We propose an innovative statistical method, called Ordinal Mixed-Effect Random Forest (OMERF), that extends the use of random forest
to the analysis of hierarchical data and ordinal responses.
%{An innovative statistical method called Ordinal Mixed-Effect Random Forest (OMERF) is proposed. It extends the use of random forest to the analysis of hierarchical data to model ordinal categorical responses.}
The model preserves the flexibility and ability of modeling complex patterns of both categorical and continuous variables, typical of tree-based ensemble methods, and,
at the same time, takes into account the %nested 
structure of hierarchical data, modeling the dependence structure induced by the grouping and allowing statistical inference at all data levels.
%In this study, 
A simulation study is conducted to validate the performance of the proposed method and to compare it to the one of other state-of-the art models. %{OMERF against existing %classical  models.}
The application of OMERF is exemplified in a case study focusing on predicting students performances using data from the Programme for International Student Assessment (PISA) 2022. %modeling students at risk of failure from a prestigious high school in Milan, and generally on predicting their academic progress. 
The model identifies discriminating student characteristics and estimates the school-effect. %{of each class to which students belong.}
\end{abstract} %%%%%%%%%

\textbf{Keywords}: Mixed-effects  models, Ordinal models, Tree-based methods, Random forest, Learning analytics.
%how to cite Bibliography
%\noindent \lipsum[1] \citet{1}

$\,$
%-----------------------------------------------------------------------------
% INTRODUCTION
%-----------------------------------------------------------------------------
\section{Introduction}
%\chapter{Introduction}
\label{sec:introduction}
Ordinal-scale observations %{are data that} 
can be categorized into a finite and ordered set of discrete categories. %{However, the} 
Distances between categories can be uneven or unknown.
These scales are typically constructed by condensing continuous variables into a set of distinct categories. An ordinal variable can be considered quantitative %{data because} 
since each level of the scale denotes a greater or lesser magnitude of a particular characteristic compared to another level. \citep{agresti2010analysis}. 
%\textit{"In all fields, ordinal scales result when inherently continuous variables are measured or summarized by researchers by collapsing the possible values into a set of categories. [\dots]
%An ordinal variable is quantitative, however, in the sense that each level on its scale refers to a greater or smaller magnitude of a certain characteristic than another level.Such variables are of quite a different nature than qualitative variables, which are measured on a nominal scale and have categories that do not relate to different magnitudes of a characteristic"}.
The growing importance of ordinal categorical data shows a clear positive trend, driven by the increasing use of surveys and tests \citep{yang2020algorithm}.
This type of data is useful to collect detailed information, especially in fields like market research, public opinion analysis, and healthcare, where assessing opinions, preferences, and responses is extremely important. 
Order becomes relevant when the categories take on meanings related to strength of opinion, agreement (as in a Likert-type response) or frequency.
An explanatory example is the case where a response variable takes on four possible values: (1) strongly disagree, (2) disagree, (4) agree, (5) strongly agree. There is a natural order in the response possibilities. 

As data collection expands into various areas, there is a bigger need to model ordered data.
Ordinal classification, often known as ordinal regression \citep{mccullagh1980regression}, represents a type of multi-class classification where there is an inherent ordering relationship between the classes, but where there is not a meaningful numeric difference between them.
%This type of problem occurs frequently in human created scales, which cover many domains from product reviews to medical diagnosis.
%{In this context,} 
This paper proposes an innovative tool for ordinal classification that extends the use of random forest \citep{breiman2001random} to the case of ordinal responses and hierarchical observations \citep{pinheiro2006mixed}.
%{It is an extension of the standard random forest method \citep{breiman2001random} to ordinal data, that while investigating the learning process, takes also into account the underlying nested structure using mixed-effects models \citep{pinheiro2006mixed}.}
The proposed method, called Ordinal Mixed-Effects Random Forest (OMERF), fits into the context of tree-based mixed-effects models \citep{hajjem2011mixed, sela2012re, hajjem2014mixed, hajjem2017generalized, pellagatti2021generalized}.
%\GB{in effetti aggregated models era troppo generico, è riferito poi a tutti i modelli citati nella literature review. Riferendoci a tree-based mixed-effects models mi sembra più chiaro. Non aggiungerei le reference considerando che sono quelle dei modelli che poi spieghiamo nel capitolo dopo. Secondo voi così va bene?} \LR{Penso sia il caso di citare qualcosa comunque, io pensavo al paper gmerf che ha la review della letteratura su questi metodi.}
In particular, the algorithm we implement disentangles the estimations of fixed and random effects, by iteratively fitting (i) a random forest \citep{breiman2001random}, ignoring the grouped data structure, and (ii) a cumulative mixed-effects model \citep{grilli2011multilevel, tutz1996random}, based on the residuals of the random forest structure. A final mixed-effects random forest is reported. To the best of our knowledge, this is the first time that a multilevel random forest for ordinal response is constructed.

The paper is structured as follows. Section \ref{sec:literature_rev} conducts a review of the literature related to analogous methods. Section \ref{sec:methods} articulates the OMERF method, outlining its theoretical foundations and its implementation.  Section \ref{sec:sim} reports a simulation study in which we test OMERF and compare it to other counterpart methods. Section \ref{sec:case} delves into a real-world case study, in which the efficacy of the proposed method is proved through the application to data from the Programme for International Student Assessment (PISA) 2022. %(\cite{organisation2023pisa})
The aim is the model students' mathematical performance levels, considering students' nested structure within schools. %\CM{QUI CAMBIEREI LA FRASE VISTO CHE NON LO CITIAMO PIù PRIMA E DAREU QUALCHE DETTAGLIO (UN PAIO DI RIGHE) SUL TIPO DI APPLICAZIONE}. 
Ultimately,  Section \ref{sec:conclusions} is dedicated to highlighting conclusions and fostering a discussion.

All the analysis are performed using R software \citep{rlanguage} and all the R codes for the OMERF algorithm and for both simulation and case study are available in the following Github repository: \url{https://github.com/giuliabergonzoli/OMERF}.

\section{Literature review}
\label{sec:literature_rev}
The method proposed in this study takes inspiration from tree-based mixed-effects models,
% \CM{QUESTI AGGREGATED VENGONO GIà CITATI MA IO PER ESEMPIO NON NE SONO DEL TUTTO AL CORRENTE, AGGIUNGEREI UNA REFERENZA (ANCHE NELLA INTRO QUANDO VENGONO CITATI.}
%{These methods refer to that type of algorithms} 
which focus on incorporating tree-based methods \citep{breiman2017classification} and their ensembles \citep{breiman2001random} into the framework of mixed-effects models \citep{pinheiro2006mixed}.
Tree-based methods are %{increasingly} 
popular for their ability to capture complex and nonlinear relationships. When declined into the mixed-effects models framework, they can handle both nested and longitudinal data.
%{this combination with mixed models allows their extension to the analysis of both nested and longitudinal data.} 
The first are data that present a hierarchical %{multilevel} 
structure, the second refer to the situation where repeated observations are available for each sampled object.
Nested data are not independent and identically distributed (\textit{i.i.d.}) as assumed in classical regression and classification models, but their distribution depends on their grouping structure. Analysing and disentangling the effects associated to each level of the hierarchy enables a deeper understanding and investigation of the regression dynamics%{latent structure present at the highest level of the hierarchy}
, thereby increasing the comprehension of the phenomenon described by the data. It is important to account for this structure, as it can provide significant insights that might otherwise be neglected, enabling the quantification of the portion of variability in the response variable that is attributable to each level of grouping.

Tree-based mixed-effects models developed in statistical literature can be categorized into two groups: the first focuses on Gaussian responses, making it unsuitable for classification tasks, while the second group extends its applicability to non-Gaussian responses and is suitable for addressing classification problems.
The works in \citep{sela2012re,hajjem2011mixed,hajjem2014mixed} pertains to the first collection of models.
In \citep{sela2012re}, the Random Effects Expectation-Maximization (RE-EM) tree is implemented; while in \citep{hajjem2011mixed}, the authors propose the Mixed-Effect Regression Tree (MERT) model.
They both are extensions of conventional regression trees to account for clustered and longitudinal data, substituting the fixed effect component of a linear mixed-effects model with a tree structure. 
Then, trying to improve prediction accuracy, a method known as Mixed-Effects Random Forest (MERF), in which the fixed effect component is modelled by a random forest, is introduced in \citep{hajjem2014mixed}. With regard to the extensions to other types of %{outcomes} 
responses, in \citep{hajjem2017generalized}, the authors propose %{linear structure used to model the fixed effects component in a generalized linear mixed model's predictor is replaced with a regression tree structure, obtaining} 
the Generalized Mixed-Effects Regression Tree (GMERT), which basically extends the MERT approach to non-Gaussian responses.
%{Another expansion of a classification problem} 
An alternative, proposed in \citep{fontana2021performing}, is the Generalized Mixed-Effects Tree (GMET), that follows a three-step procedure: first, the random-effects are initialised to zero and the systematic component is estimated through a generalized linear model; then, a regression tree is built using the estimated systematic component as dependent variable and, finally, a mixed-effects model is fitted to estimate the random-effects part, using the estimated tree as offset.
In \citep{fokkema2018detecting}, the authors propose an algorithm known as Generalized Linear Mixed-effects Model tree (GLMM tree), which iteratively refines the estimates of a generalized linear model tree and a mixed-effects model until convergence is achieved.
Lastly, in \citep{speiser2020bimm}, a decision tree method for modeling clustered and longitudinal binary outcomes within a Bayesian framework, called Binary Mixed Model tree (BiMM tree), is introduced.
A step forward in the field of tree-based aggregated models has been done in \citep{pellagatti2021generalized}, in which the authors implement the Generalized Mixed-Effects Random Forest (GMERF), thus extending for the first time random forest, and not only simple trees, to deal with hierarchical data, both for regression and classification (for any response variable in the exponential family).

Contributing to this branch of the literature, the current research work proposes a novel method called Ordinal Mixed-Effects Random Forest (OMERF), that is inspired by the GMERF model, but extends the multilevel random forest approach to deal with ordinal data, namely, with ordinal regression problems.

Concerning the statistical literature about ordinal data, one of the early contributions to classification techniques for ordinal data can be found in \citep{mccullagh1980regression}, where a regression model is introduced. Moving to a multilevel setting, the random effects cumulative model, described in \citep{tutz1996random}, addresses ordinal regression models as special cases of multivariate generalized linear models, adjusting them to include random effects in the linear predictor.
With regard to the implementation of nonlinear ordinal models, some attempts can be found in the literature.
The work in \citep{tutz2003generalized} proposes an extension of ordinal regression through the generalization of the additive model by incorporating nonparametric terms; in \citep{shashua2002ranking} the authors introduce a generalised formulation for the support vector machine for ordinal data. The ordinal random forest method is presented in \citep{hornung2020ordinal}, which is a random forest-based prediction method for ordinal response variables. Finally, in \citep{tutz2022ordinal}, an extension with score-free recursive partitioning and ensembles that include parametric models is proposed.

%Although linear models for nested data and ordinal responses, as well as tree-based methods for ordinal data, have been proposed in the literature, we are not aware of tree-based methods being proposed for nested data and ordinal responses. 
Within this literature, OMERF, combining random effects
cumulative models \citep{tutz1996random} with the ordinal random forest \citep{hornung2020ordinal}, is the first tree-based method for nested data and ordinal responses and paves the way for a new class of models of increasing interest.

%{Nonetheless, to the best of authors' knowledge, this is the first time that random forest is extended to deal with both the ordinal classification setting and nested structures, implementing an ordinal model which is able to capture complex nonlinear data relationships, and at the same time to inspect and understand the multilevel nature of data.}

%-----------------------------------------------------------------------------
% EQUATIONS
%-----------------------------------------------------------------------------
\section{Model and Methods}
\label{sec:methods}
In this section, a concise overview of cumulative link models for ordinal data (%Section \ref{sec:clm} and 
Section \ref{sec:clmm}) and random forest (Section \ref{sec:rf}) is provided.
This will serve to set the notation and to understand the OMERF method, detailed in Section \ref{sec:omerf}. %Finally, Section \ref{sec:gof} is devoted to inspect different performance measures to evaluate the Goodness of Fit of ordinal models: these measures will serve as assessment criteria for the compared models.

\subsection{Background and state of the art models}

\subsubsection{Cumulative link mixed models}
\label{sec:clmm}

%These models applied to ordinal responses are advantageous in the fact that there is no need to define interval-scale assumptions about distances between response categories.
Ordinal observations can be expressed through a random variable, $Y_{j}$, which takes on the value $c$ when the $j$-th ordinal observation assumes the $c$-th category. The support of $Y_{j}$ includes the integer values from 1 to $C$, where $C \geq 2$.
Cumulative Link Models (CLM) \citep{agresti2010analysis,ananth1997regression,mccullagh1980regression} deal with observations on an ordinal scale and apply an arbitrary link function to link the cumulative probabilities to a linear predictor.
Cumulative Link Mixed Models (CLMM) \citep{grilli2011multilevel,tutz1996random} extends CLMs to deal with nested observations, by including normally distributed random effects. %{are the multilevel extension of cumulative regression models with normally distributed random effects for ordinal responses.}
These models are suited to handle data with a hierarchical structure. Given $\bm{y}_{i}$ = $y_{i1}, \dots, y_{in_{i}}$ the $n_{i}$-dimensional response vector for observations in the $i$-th group, 
%{The key part of this standard assumption is the exogeneity, namely the mean of the random effects does not depend on the fixed covariates: $\mathbb{E}(\bm{b}_{i}|\bm{x}_{ij}:j=1,\dots,n_{i})=0$.}
the configuration of a CLMM with a random intercept and $Q$ random slopes in a two-level hierarchy can be written as:
\begin{equation}
    \label{clmm}
    \begin{aligned}
        \eta_{ijc} = g(\gamma_{ijc}) &= \theta_{c} - (\bm{x}_{ij}^T \bm{\beta} + \bm{z}_{ij}^T \bm{b}_{i}),\qquad c=1,\dots,C-1 \\
        \bm{b}_{i} &\sim \mathcal{N}_Q(\bm{0},\bm{\Sigma_{b}})
    \end{aligned}
\end{equation}
where $i=1,\dots,I$ is the level 2 (group) index and $j=1,\dots,n_{i}$ (with $\sum_{i=1}^{I} n_{i}=J$) is the nested level 1 index. $\gamma_{ijc} = \mathbb{P}(y_{ij} \leq c) = \pi_{ij1} + \dots + \pi_{ijc}$ (with $\sum_{c=1}^{C} \pi_{ijc}=1$ and $\pi_{ijc}=\mathbb{P}(y_{ij} = c)$) is the cumulative probability up to the \textit{c}-th category for unit \textit{j} in group \textit{i}. $\eta_{ijc}$ is the linear predictor, $\bm{x}_{ij}$ is a $P$-dimensional vector of predictors, to which corresponds the fixed effects vector $\bm{\beta}$, associated to the entire population, and $\bm{z}_{ij}$ is a $(Q+1)$-dimensional vector of predictors, to which correspond group-specific parameters $\bm{b}_{i}$. The elements $y_{ij}$ are supposed to be conditionally independent on the random effects $\bm{b}_{i}$.
Moreover, random effects $\bm{b}_{i}$, conditionally on the fixed effects, are assumed to be independent and identically distributed with zero mean and a common group covariance matrix $\bm{\Sigma_{b}}$. Finally, $g$ is a monotonic, differentiable link function and $\theta_{c}$ are the strictly ordered thresholds (also known as cut-points or intercepts):
\begin{gather*}
    \begin{aligned}
        -\infty \equiv \theta_{0} \leq \theta_{1} \leq \dots \leq \theta_{C-1} \leq \theta_{C} \equiv \infty.
    \end{aligned}
\end{gather*}
%Since the overall intercept is fixed to zero, if it is the case of a single random intercept and no random slopes, the random effect $b_{0i}$ representing unobserved factors at the cluster level, can be seen as a random shift of the threshold parameters $\theta_{c}$ so that the set of thresholds of cluster $i$ is $\theta_{c}-b_{0i}$.

In CLMMs, the ordinal response variable $y_{ij}$ with $C$ categories is assumed to be generated by a latent continuous variable $y^*_{ij}$  with a set of $C-1$ thresholds $\theta^*_{c}$ such that $y_{ij}=c$ $\iff$  $\theta^*_{c-1} \leq y^*_{ij} \leq \theta^*_{c}$.
The latent continuous variable is modelled as:
\begin{equation}
    \label{cont_clmm}
    \begin{aligned}
        y^*_{ij} = \bm{x}^{T}_{ij} \bm{\beta^*} + \bm{z}_{ij}^T \bm{b}^*_{i} + \epsilon^*_{ij}
    \end{aligned}
\end{equation}
where $\epsilon^*_{ij}$ is a level 1 error with standard deviation $\sigma_{\epsilon^*}$.

Therefore, the cumulative probabilities are $\gamma_{ijc} = \mathbb{P}(y_{ij} \leq c) = \mathbb{P}(y^*_{ij} \leq \theta^*_{c}) = \mathbb{P}(\epsilon^*_{ij} \leq \theta^*_{c} - \bm{x}^{T}_{ij} \bm{\beta^*} - \bm{z}_{ij}^T \bm{b}^*_{i}) = g^{-1}(\theta_{c} - \bm{x}^{T}_{ij} \bm{\beta} - \bm{z}_{ij}^T \bm{b}_{i})$.
The underlying linear model \eqref{cont_clmm} with thresholds $\theta^*_{c}$ and level 1 error $\epsilon^*_{ij}$ having distribution function $g^{-1}$ is equivalent to the cumulative model \eqref{clmm} with link function $g$.
%A generic parameter $\alpha$ of the cumulative model can be retrieved as $\alpha = \alpha^* \frac{\sigma_{g}}{\sigma_{\epsilon^*}}$, where $\sigma_{g}$ is the standard deviation of the distribution associated to the link function $g$. \CM{mi sono persa qui con alpha, cos'è?}

In the context of CLMMs, model parameters are typically estimated using maximum likelihood methods, employing techniques such as Adaptive Gaussian Quadrature, Gauss-Hermite Quadrature, or Laplace approximation to the likelihood function.
A Newton-Raphson algorithm updates the conditional modes of the random effects for the subsequent approximations and finally a nonlinear optimization is performed over the fixed parameter set to get the Maximum Likelihood Estimation.

\subsubsection{Random forest}
\label{sec:rf}
Random forests (RFs) are a powerful machine learning algorithm that combines the predictive power of multiple decision trees to make accurate and robust predictions.
In particular, random forest for regression \citep{breiman2001random,james2013introduction} are tree-based ensemble methods formed by growing regression trees such that the tree predictor $f(\bm{x})$ takes on numerical values, and the random forest predictor is formed by taking the average over $K$ of these trees $f_k(\bm{x})$, $k=1,\dots,K$.

The extension for ordinal responses, namely ordinal forests (OF) \citep{hornung2020ordinal}, involves the construction of regression forests in which traditional class values are replaced by score values. These score values are optimized to improve out-of-bag (OOB) prediction performance, evaluated against a designated measure known as the performance function. 
This model is based on the assumption that exists a continuous variable, denoted as $y^*$, underlying the observed ordinal variable $y$, known or unknown, indicating the ordinal variable's values. Specifically, the relationship dictates that as the value of $y^*$ increases for an observation, so does the corresponding class of the ordinal response variable. Class widths are the widths of $J$ adjacent intervals and can vary between class and class.

% \CM{AGGIUNGERE DUE MICRO COSE SULLA WIDTH E LATENT Y CONTINUA DAL PAPER}

\subsection{Ordinal mixed-effects random forest}
\label{sec:omerf}
The proposed statistical method, called Ordinal Mixed-Effects Random Forest (OMERF), extends the use of random forest to the analysis of hierarchical data, for categorical ordinal response variables.
It models the fixed effects through a random forest, combining them to the random effects obtained using a CLMM, in order to take into account both possible complex functional forms in the fixed effect component and the nested structure of data.
The model can be formulated as follow:
\begin{equation}
    \label{eq:omerf}
    \begin{aligned}
        \eta_{ijc}  = g(\gamma_{ijc}) &= \theta_{c} - (f(\bm{x}_{ij}) + \bm{z}_{ij}^T \bm{b}_{i})\\
        g(\gamma_{ijc}) = logi&t(\gamma_{ijc}) = \ln(\frac{\gamma_{ijc}}{\gamma_{ijc}-1}) \\
        \gamma_{ijc} &= \mathbb{P}(y_{ij} \leq c) \\
        \bm{b}_{i} &\sim \mathcal{N}_Q(\bm{0},\bm{\Sigma_{b}}) \\
        j=1,\dots,n_{i} \qquad i&=1,\dots,I \qquad c=1,\dots,C-1
    \end{aligned}
\end{equation}
where $f(\bm{x}_{ij})$ is the unknown and nonlinear structure estimated through the random forest,
$\gamma_{ijc}$ are cumulative probabilities, $\pi_{ijc} = \mathbb{P}(y_{ij}=c) = \mathbb{P}(y_{ij} \leq c) - \mathbb{P}(y_{ij} \leq c-1) = logit^{-1}(\theta_{c} - (f(\bm{x}_{ij}) + \bm{z}_{ij}^T \bm{b}_{i})) - logit^{-1}(\theta_{c-1} - (f(\bm{x}_{ij}) + \bm{z}_{ij}^T \bm{b}_{i}))$ is the probability that the $j$-th observation, within the $i$-th group, falls in the $c$-th category. %{and $\eta_{ijc}$ is the linear predictor}.
Similarly to CLMM model, OMERF model assumes that the random effects $\bm{b}_i$ and $\bm{b}_{i'}$ are independent for $i \neq i'$.
%The fixed component is modelled by a nonparametric RF model that concerns the entire population, while random effects are identified through group-specific parameters.
% \CM{FORSE NEI METODI VALE LA PENA METTERE UNA FRASE IN CUI DICIAMO CHE LA STIMA DELLA FIXED COMPONENT DI FATTO E' UN RF OBJECT (SE GIà NON LO DICIAMO ESPLICITAMENTE) E QUINDI TUTTA LA PARTE DI INTERPRET (PARTIAL PLOTS E IMPOPLOT ECC) VIENE COME SEMPRE}
The fixed component is described by a RF object and, consequently, its exploration relies on  familiar tools such as partial plots and variable importance plots.
%\GB{Therefore, the estimation of the fixed component is indeed an RF object, and consequently, all the interpretation of the results refers to this type of model, exploiting familiar tools such as partial plots and variable importance plots.}

The OMERF algorithm is inspired by the one proposed in \citep{pellagatti2021generalized} and estimates fixed and random effects by following an iterative procedure in which the two components are estimated separately.  %one must separate the estimation of the fixed and random effects components and alternate between them until convergence.
To perform this estimation, it can be observed that if the random effects were known in advance, a RF could be fitted to estimate the fixed part $f(\bm{x}_{ij})$ by using $\eta_{ijc} + \bm{z}^T_{ij}\times \bm{b}_i$ as dependent variable.
Similarly, if the population-level effects were known, the random effects could be estimated using a CLMM with the response corresponding to $\eta_{ijc}$ and using $f(\bm{x}_{ij})$ as an offset of the model.
Since neither of them is known, an iterative approach that alternates between estimating the RF for the fixed component and estimating the CLMM for the random one is employed.
Convergence is considered achieved when the difference between the random effects estimates in two consecutive iterations is less than a predetermined tolerance.

% \CM{rigiro l'ordine di queste info, prima l'algoritmo e i riferimenti ai pacchetti, poi da ultimo il punto delicato dell'inizializzazione}

The pseudo-code outlining this estimation process is provided in Algorithm \ref{alg:alg_omerf}. The random forest model is constructed using the R package $randomForest$ \citep{RF}, which implements the original algorithm described in \citep{breiman2001random}. Meanwhile, the CLMM is built using the $clmm$ function from the R package $ordinal$ \citep{ordinal}.
The CLMM model allows for different offsets in the formula and scale effects, which are considered as components of the linear predictor that are known in advance and thus  they require no parameter to be estimated from the data. The implemented method in particular will make use of an offset modelled as:
\begin{equation}
    \label{scale_eff}
    \begin{aligned}
        \eta_{ijc}  = g(\gamma_{ijc}) = \theta_{c} - \bm{z}_{ij}^T \bm{b}_{i} - offset_{ij},\\  j=1,\dots,n_{i} \hspace{0.5cm}  i=1,\dots,I \hspace{0.5cm} c=1,\dots,C-1
    \end{aligned}
\end{equation}
where $offset_{ij}$ =  $f(\bm{x}_{ij})$, i.e., the random forest estimates of the fixed component.

Addressing the initialization of the unknown systematic component $\eta_{ijc}$ represents a delicate challenge, as it cannot be directly inferred from the data. To tackle this issue, we utilize a traditional Ordinal Forest model incorporating covariates of the fixed component as predictors and the vector with ordinal categorical responses {$y_{ij}$ as target.
This model is used to estimate the cumulative probabilities $\gamma_{ijc}$ and, then, the inverse link function $g^{-1}$ is applied to initialize $\eta_{ijc} = g^{-1}(\gamma_{ijc})$. The implementation of the ordinal forest can be found in the \(ordinalForest\) R package \citep{hornung2020ordinal}.
% \CM{AGGIUNGERE QUALCHE DETTAGLIO SU COME VIENE STIMATA LA ETA E IL RIFERIMENTO AL PACCHETTO.} 
This approach represents an enhancement compared to similar methods like GMERF \citep{pellagatti2021generalized}, which employs a GLM for initializing $\eta_{ij}$, thereby missing out on the benefits of non-parametric methods, at the step 0 of the algorithm. Specifically, using a GLM for initialization fails to capture potentially nonlinear trends and interactions, a capability that OMERF aims to achieve.
%A delicate issue concerns the initialization of the systematic component $\eta_{ijc}$ that is unknown and cannot be directly inferred from the data. To tackle this aspect, we employ a conventional Ordinal Forest model that includes covariates of the fixed component as predictors. AGGIUNGERE QUALCHE DETTAGLIO SU COME VIENE STIMATA LA ETA E IL RIFERIMENTO AL PACCHETTO.  Estimating the systematic component $\eta_{ijc}$ by means of Ordinal Forest represents an improvement with respect to counterpart methods such as GMERF \citep{pellagatti2021generalized}, that uses a GLM to initialize $\eta_{ij}$, losing the advantages of non parametric methods. Indeed, using a generalized linear model to initialize the systematic component does not allow capturing potentially nonlinear trends and interactions, which, instead, OMERF aims to capture. induces all limitations of parametric methods

Once OMERF has been fitted, to make predictions for a new observation $[\bm{x}_{ij};\bm{z}_{ij}]$, the following formula is employed: $\hat{\eta}_{ijc} = \hat{\theta}_{c} - (\hat{f}(\bm{x}_{ij}) + \bm{z}_{ij}^T \hat{b}_{i})$.
Here, $\hat{f}$ represents the random forest model estimated by the algorithm, $\hat{b}_i$ is the vector of random effects %{coefficients}
associated to the $i$-th group, and $\hat{\theta}_{c}$ is the threshold associated to each predicted category $c$.

\begin{algorithm}[H]
\caption{OMERF}
\label{alg:alg_omerf}
\scriptsize
\begin{algorithmic}[1]
\STATE \textbf{Input}:
    \begin{itemize}[label={}, leftmargin=*,noitemsep,nolistsep]
        \item {$y-$} vector with ordinal categorical responses {$y_{ij}$}
        \item {$cov-$} data frame with all covariates
        \item {$group-$} vector with the grouping variable for each observation
        \item {$xnam-$} vector with names of the covariates to be used as fixed effects
        \item {$znam-$} vector with names of the covariates to be used as random effects
        \item {$b_{0}-$} optional matrix of initial values for each {$\underline{b}_{i}$}
        \item {$toll-$} threshold to decide whether our estimation converged or not (default value of 0.05)
        \item {$itmax-$} maximum number of iterations (default value of 100)
    \end{itemize}
\STATE {$Z \leftarrow$} (1;{$cov[znam]$}): it includes also the random intercept
\STATE Initialize {$b$} to a matrix of zero (if {$b_{0}$} is not given): each column {$b[,i]$} of {$b$} will be the {$i$}-th random coefficients {$\underline{b}_{i}$}
\STATE {$all.b[0]=b$}
\STATE fit a Ordinal Forest model using {$y$} as response and {$cov$} as matrix of covariates
\STATE {$eta \leftarrow$} estimated {$\eta_{ijc}$} by the Ordinal Forest model
\STATE {$it \leftarrow$} 1
\WHILE {$it < itmax$ \AND \NOT {$conv$}}
    \STATE {{$targ \leftarrow eta + Z \times b$}}
    \STATE fit a random forest model using {$targ$} as target and {$cov$} as predictor matrix
    \STATE {$fx \leftarrow$} fitted values of the forest model
    \STATE fit the CLMM model {$\eta_{ijc} = \theta_{c} - \underline{z}_{ij}^T \underline{b}_{i} - offset_{ij}$}, with {$offset_{ij} =  fx$}
    \STATE {$all.b[it]=b \leftarrow$} the estimated {$b$} fromm CLMM model
    \STATE {$M \leftarrow$} max({$abs(b-all.b[it-1])$})
    \STATE {$(i,j) \leftarrow$} argmax({$abs(b-all.b[it-1])$})
    \STATE {$tr \leftarrow M/all.b[it-1](i,j)$}
    \IF {$tr < toll$}
        \STATE {$conv \leftarrow$} \TRUE
    \ELSE
        \STATE {$conv \leftarrow$} \FALSE
    \ENDIF
    \STATE {$it++$}
\ENDWHILE

\IF {\NOT $conv$}
    \STATE give a warning
\ENDIF
\STATE \textbf{Output}:
    \begin{itemize}[label={}, leftmargin=*,noitemsep,nolistsep]
        \item {$clmm.model-$} the final CLMM model fitted
        \item {$forest.model-$} the final forest model fitted
        \item {$b-$} the final estimation of the random coefficients
        \item {$it-$} the number of iterations
    \end{itemize}

\end{algorithmic}
\end{algorithm} 

\section{Simulation study}
\label{sec:sim}
In this section, we conduct a simulation study to test OMERF and compare it with similar classification methods under various simulation settings.
%{In this section, a simulation study is reported to compare the performance of OMERF with that of similar classification methods on various simulated datasets.}
In Section \ref{sec:simdesign}, we describe the design of the %{created datasets} 
data generating process (DGP), while, in Section \ref{sec:simres}, results are analysed, in order to highlight strengths and weaknesses of OMERF.

\subsection{Simulation design}
\label{sec:simdesign}

To sample ordered categorical data, we make use of the function \(genOrdCat()\) from the R package \(simstudy\) \citep{simstudy}.
This function takes as input an underlying (continuous) latent process \(w_{ij}\) 
% \CM{HO AGGIUNTO IL PEDICE IJ ALLA W, CORRETTO?}
as the basis for data generation.
Assuming that probabilities are determined by segments of a logistic distribution, it defines the ordinal mechanism
using thresholds along the support of the distribution. %{For example,} 
In case of \(C\) possible responses, there will be \(C-1\) thresholds.
The area under the logistic density curve of each of the \(C\) regions defined by those thresholds %{(there will be \(k\) distinct regions)} 
represents the probability of each possible response tied to that region.

 In our simulation, we set \(C\) = 3 and we generate the underlying latent process as:
\begin{equation}
    \label{eq:simstudy1}
    \begin{aligned}
        w_{ij} =f(\bm{x}_{ij}) + \sum_{q=0}^{Q} z_{qij}^T b_{qi} \qquad j=1,\dots,n_{i} \qquad i&=1,\dots,I 
    \end{aligned}
\end{equation}

where \(f\) is the fixed component functional form which takes in input the \(P\)-dimensional vector of fixed effects covariates $\bm{x}_{ij}$ and \(\sum_{q=0}^{Q} z_{qij}^T b_{qi}\) is the random component. The ordinal response $Y$ is generated from $w_{ij}$ by using the function \(genOrdCat()\) assuming balanced categories. 

Regarding the fixed effects part, we use a variation of the simulation design proposed in \citep{pellagatti2021generalized}.
The design for \(f\) incorporates both a linear component and a tree-like component, along with interactions among the covariates.
This approach allows to simulate scenarios with highly diverse structure, which will challenge the flexibility and adaptability of our method.

Specifically, $P=7$ fixed effects covariates are taken into account and \(f\) is modelled as follows:
% \CM{VISTO CHE DOPO LE DISTRIBUZIONI X1 X2 ECC COMPAIONO IN MAIUSCOLO, FORSE LE SCRIVEREI IN MAIUSCOLO ANCHE QUI NELL'EQ 12 E NEL TESTO, ALBERO ECC}
\begin{equation}
    \label{eq:simstudy2}
    \begin{aligned}
        f(X_{1}, \dots,X_{7}) = \alpha(3+7X_{1}^2-5X_{2}+X_{2}X_{3}^2) + \beta tree(X_{4},X_{5},X_{6})
    \end{aligned}
\end{equation}

\noindent
where \(\alpha\) and \(\beta\) represent two design parameters employed to regulate the importance given to the linear and tree-based components in the various DGPs.
The function \(tree(X_{4}, X_{5}, X_{6})\) follows the tree like stucture outlined in Figure \ref{fig:tree}.
The variable \(X_{7}\) is included even though it is not significant, in order to assess whether the algorithm is influenced by it. Indeed, while all of the seven variables are
being used as predictors in the compared models, only the first six of them are actually used to generate \(f\).

The seven variables are generated randomly in accordance with the following distributions:
\(X_{1},X_{2},X_{3}\sim N(0,1); X_{4}\sim U(-3,3); X_{5}\sim U(-6,6); X_{6}\sim U(-5,5); X_{7}\sim U(-4,4)\).

\begin{figure}[H]
    \centering
    \includegraphics[width=0.65\textwidth]{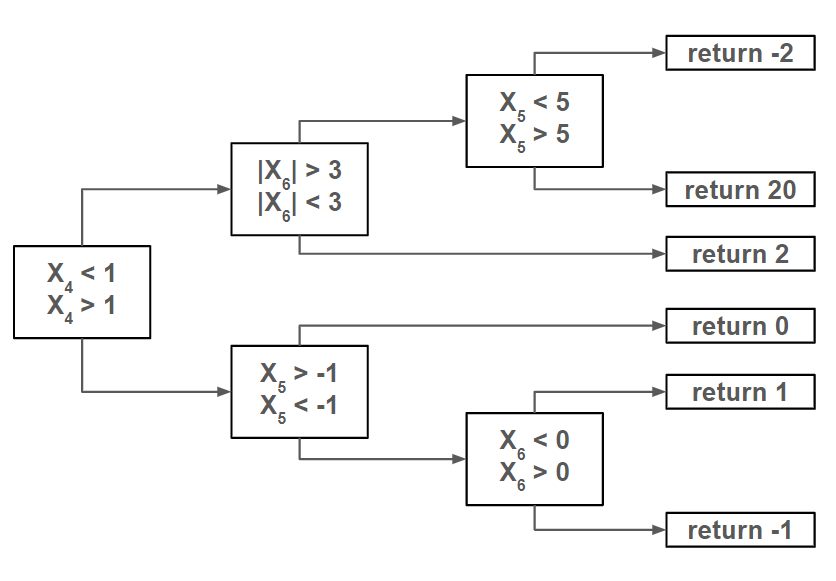}
    \caption{Tree-like structure \(tree(X_{4},X_{5},X_{6})\) of the fixed effects part in Equation \ref{eq:simstudy2}.}
    \label{fig:tree}
\end{figure}

The random effects are drawn from a normal distribution for two distinct scenarios:
\begin{itemize}
    \item \(Random \quad intercept \quad only: \sum_{q=0}^{Q} z_{qij}^T b_{qi} = b_{0i} \sim \mathcal{N}(0, \sigma^2_1)\), where \(\sigma^2_1\) is a design parameter which gives the possibility to change the variability of the effects in the different simulations.
    Indeed, within each scenario, two specifications (low and high) for the parameter \(\sigma^2_1\) are considered in order to account for different levels of magnitude of the between-group variability;
    \item \(Random \quad intercept \quad and \quad slope: \sum_{q=0}^{Q} z_{qij}^T b_{qi} = b_{0i} + x_{1ij}^T b_{1i}\), where \(X_{1}\) has been previously defined and \(\bm{b}_i \sim \mathcal{N}_2(0, \Sigma)\), with \(\Sigma=diag(\sigma^2_1;\sigma^2_2)\).
    It is important to note that the random effects \(b_{0i}\) and \(b_{1i}\) are treated as independent for any given value of \(i\), and, as in the previous scenario, \(\sigma^2_1\) and \(\sigma^2_2\) regulate the random effects variance.
    In this setting, the covariate \(X_1\) is not assumed to have a fixed effect that applies uniformly to the entirety of observations. Instead, its association to the response is considered group-specific, meaning that \(X_1\) is assumed to have different effects across observations belonging to distinct groups.
\end{itemize}

A two-level data structure of \(I = 10\) groups with \(n_{i} = 100\) observations each is simulated, for a total number of 1000 units.
Note that for simplicity, an equal number of observations for each group is taken to ensure a balanced dataset. However, the model is, of course, capable of handling datasets with varying group sizes.

In addition to this simulation design, a full linear DGP is implemented in order to observe how OMERF performs in a parametric context. In this case, the fixed component includes only the first three variables previously described and is defined as: %{the first three covariates generated before are combined as follows}:
\begin{equation}
    \label{eq:simstudy3}
    \begin{aligned}
        f(X_{1}, \dots,X_{3}) = 3+7X_{1}-5X_{2}+X_{2}X_{3}.
    \end{aligned}
\end{equation}
Only the case of a single, normally distributed random intercept is considered: \(b_{0i} \sim \mathcal{N}(0, \sigma^2_1)\).

The parameters listed in Table \ref{table:DGPs} are chosen in order to obtain balanced datasets regarding the classes of the ordinal response.
Moreover, several combinations of parameters are tested to observe how the model reacts in the case of a more polynomial or more tree-like systematic component and with the variance related to random effects being higher or lower.
Thus, a total of 10 DGPs summarised in Table \ref{table:DGPs} are obtained. The main objective of the designed DGPs is to show how our algorithm is able to capture %{the variability of the groups} 
the heterogeneity across groups and the structure of the fixed component.

OMERF performance is evaluated in comparison to other three models: 
\begin{itemize}
    \item CLM and CLMM, from the \(ordinal\) R package \citep{ordinal}, which are expected to perform better in a full linear context, but worse in cases of more complex structures, and CLM is expected not to grasp the hierarchical structure of data;
    \item Ordinal random forest, from the \(ordinalForest\) R package \citep{hornung2020ordinal}, which, as typical of ensemble tree-based models, is able to capture nonlinear relationships, but not to catch the nested structure.
\end{itemize}
These methods are chosen in order to, at least to some extent, include a range of both cumulative link and tree-based models.

% \CM{IO SPOSTEREI QUI LA SEZIONE SUI GOF, DICENDO CHE LE PERFORMANCE DI QUESTI METODI SONO CONFRONTATE IN TERMINI DI QUESTE METRICHE.}
The choice of appropriate performance metrics for ordinal models is not straightforward and is an area of research still not widely explored \citep{de2021comparison}. Therefore, we compare the performance of the tested methods by employing multiple godness of fit (gof) metrics. We consider the accuracy, that is designed to deal with categorical data and allows to track the percentage of correctly classified observation, but not the error severity, and the Mean Square Error (MSE), that treats the ordinal scale as real number \cite{gaudette2009evaluation}. Furthermore, we incorporate two indexes that assess the similarity between two classifications of the same objects by quantifying the agreement proportions between the two partitions.  These are the Adjusted Rand Index \citep{hubert1985comparing} and Cohen's kappa \citep{cohen1960coefficient}.
Both of these indices have a range from -1 to 1. Positive values in this range indicate agreement between the two sets, with 1 denoting perfect agreement. Negative values imply disagreement, and the magnitude of the negative value reflects the extent of this disagreement. A value equal to 0 suggests that the agreement is no different from what would be expected by chance. Lastly, we select among the recent developments in the field the two indexes implemented by J. S. Cardoso \citep{cardoso2011measuring} and E. Ballante \citep{ballante2022new}. These last two indexes are novel metrics specifically adapted to ordinal data classification problems, they allow values in the range [0;1] with the optimal value in 0.

\begin{table}[H]
    \centering 
    \begin{tabular}{|p{4em} | c | c | c | c | c |}
    \hline
    & \textbf{\(f(\bm{x}_{ij})\)} & \textbf{\(\alpha\)} & \textbf{\(\beta\)} & \textbf{\(\sigma^2_1\)} & \textbf{\(\sigma^2_2\)}  \\
    \hline \hline
    \textbf{DGP 1} & \(\alpha(3+7X_{1}^2-5X_{2}+X_{2}X_{3}^2) + \beta tree(X_{4},X_{5},X_{6})\) & 0.3 & 0.7 & 1 & -  \\
    \textbf{DGP 2} & \(\alpha(3+7X_{1}^2-5X_{2}+X_{2}X_{3}^2) + \beta tree(X_{4},X_{5},X_{6})\) & 0.7 & 0.3 & 1 & -  \\
    \textbf{DGP 3} & \(\alpha(3+7X_{1}^2-5X_{2}+X_{2}X_{3}^2) + \beta tree(X_{4},X_{5},X_{6})\) & 0.3 & 0.7 & 5 & -  \\
    \textbf{DGP 4} & \(\alpha(3+7X_{1}^2-5X_{2}+X_{2}X_{3}^2) + \beta tree(X_{4},X_{5},X_{6})\) & 0.7 & 0.3 & 5 & -  \\
    \textbf{DGP 5} & \(\alpha(3+7X_{1}^2-5X_{2}+X_{2}X_{3}^2) + \beta tree(X_{4},X_{5},X_{6})\) & 0.3 & 0.7 & 0.3 & 0.5  \\
    \textbf{DGP 6} & \(\alpha(3+7X_{1}^2-5X_{2}+X_{2}X_{3}^2) + \beta tree(X_{4},X_{5},X_{6})\) & 0.7 & 0.3 & 0.3 & 0.5  \\
    \textbf{DGP 7} & \(\alpha(3+7X_{1}^2-5X_{2}+X_{2}X_{3}^2) + \beta tree(X_{4},X_{5},X_{6})\) & 0.3 & 0.7 & 1 & 1  \\
    \textbf{DGP 8} & \(\alpha(3+7X_{1}^2-5X_{2}+X_{2}X_{3}^2) + \beta tree(X_{4},X_{5},X_{6})\) & 0.7 & 0.3 & 1 & 1  \\
    \textbf{DGP 9} & \(3+7X_{1}-5X_{2}+X_{2}X_{3}\) & - & - & 1 & -  \\
    \textbf{DGP 10} & \(3+7X_{1}-5X_{2}+X_{2}X_{3}\) & - & - & 5 & - \\
    \hline
    \end{tabular}
    \\[10pt]
    \caption{Simulation parameters of both fixed and random effects parts for 10 different DGPs.}
    \label{table:DGPs}
\end{table}

\subsection{Simulation results}
\label{sec:simres}
For each of the ten DGPs described in Table \ref{table:DGPs}, we simulate 100 datasets. In order to evaluate the predictive performances of the four compared models, each dataset is randomly split into training and test sets, with a ratio of 80\% for training and 20\% for testing.
%\CM{QUESTA FRASE PROBABILMENTE ANDRA' TOLTA INSERENDO PRIMA IL PARAGRAFO SUI GOF To evaluate the quality of the predictions the five performance measures presented in Section \ref{sec:gof} are computed.} 
Simulation results, in terms of mean and variance of gof metrics computed across the 100 runs, are reported in Table \ref{table:res_lin}\footnote{OMERF and the ordinal forest are run with default inputs.}.

% \CM{FRASE BLURRED From the findings reported, it can be inferred that generally, all performance measures in each simulation are consistent in the results. In particular, the performance of the created OMERF method is always good, and, even when not the best, remains mostly comparable to that of the top-performing model.}

% \CM{FORSE SPOSTEREI QUESTA FRASE PRIMA, A COMMENTO DELLA TABELLA 2 E QUI ANDREI DRITTA SUI CONFRONTI METTENDO TRA PARENTESI LA MICRO DESCRIZIONE DEL DGP. INIZIO IO POI TU FAI CHECK: DGPs 1 to 4 correspond to simulations incorporating both polynomial and tree-based structures, accounting for random intercepts but not random slopes. DGPs 1 and 2 exhibit low between-group variability, while DGPs 3 and 4 demonstrate higher variability. DGP 1 emphasizes the importance of the tree structure, whereas DGP 2 prioritizes the polynomial component.}
For DGPs from 1 to 4 (i.e., fixed component incorporating both polynomial and tree-based structures and random intercept with varying effect), OMERF consistently emerges as the optimal choice across all reported metrics. Notably, the efficacy gap between tree-based methods (ordinal forest and OMERF) and linear models (CLM and CLMM) can be appreciated both in scenarios with a preponderant tree-like structure and in the ones with a preponderant polynomial structure. The performances of the OMERF method in these scenarios highlights its ability to best capture complex relationships between target and covariates. %Particularly noteworthy is the OMERF method's superior performance in DGP 2,
% (with \(Accuracy\) = 0.8089, \(MSE\) = 0.2565, \(ARI\) = 0.5330, \(Cohen's \, k\) = 0.5066, \(Cardoso \, idx\) = 0.2836, \(Ballante \, idx\) = 0.0843) but the most substantial performance gap compared to linear models is observed in DGP 4. \GB{These are indeed DGPs with a  higher nonlinear structure, the performances of the OMERF method in these scenarios highlights its ability to best capture complex relationships between target and covariates.}

%Note that the difference in effectiveness between tree-based methods (ordianl forest and OMERF) and linear models (CLM and CLMM) is especially significant when the DGP has a tree-like structure (e.g. DGPs 1-8 in Table \ref{table:res_lin}).
%(e.g. DGPs 1,3 in Table \ref{table:res_lin}).
%On the other hand, in cases where the DGP is based on a polynomial structure (e.g. GDPs 9,10 in Table \ref{table:res_int}), the ordinal random forest and OMERF reach lower performance compared to linear models.
%models still better capture the complexity of the problem, but the difference with the basic linear models is more negligible.
%The OMERF method reaches the best performance in DGP 2 (with \(Accuracy\) = 0.8039, \(MSE\) = 0.2646, \(ARI\) = 0.5324, \(Cohen's \, k\) = 0.5092, \(Cardoso \, idx\) = 0.2918, \(Ballante \, idx\) = 0.0960), but has the largest difference in performance compared to the linear models in DGP 1, with an accuracy 0.0784 higher than CLM and 0.08 higher than CLMM.

With the introduction of a random slope (namely in the DGs 5-8), the models performance remains consistent with those analysed so far. Notably, random forest consistently outperforms linear models, although OMERF does not consistently emerge as the best performer across all indices.
Moreover, in all the simulations there appear to be no major differences between the cases with small or large variability of random effects.

On the contrary, in DGPs 9 and 10, in which the predictor is linear, CLM and CLMM tend to perform better with respect to the other models.
This result confirms that tree-based methods better capture nonlinear dependencies, but when the data structure is linear parametric, linear models are preferable, additionally yielding results that are usually more easily interpretable.

%{As for the variances in the estimations, all models estimates present small variability. Indeed, even OMERF is able to provide stable estimates, probably due to the iterative nature of the algorithm, which stabilizes the results.}

Overall, results confirm that, in a nonlinear setting, OMERF performs better with respect to CLM and CLMM, and slightly better or comparably to ordinal random forest, still having the advantage of extracting knowledge from the nested data structure.

For what concerns the estimation of the predictor, we provide an example of the OMERF output by reporting, in Figures \ref{fig:ranefDGP1} and \ref{fig:fix_cov_DGP1}, the results, in terms of fixed and random effects, in one of the runs of DGP 1\footnote{The choice of reporting results for a single run is forced by the impossibility of summarizing this type of outcome across the runs.}. Figure \ref{fig:ranefDGP1} reports the sampled (Figure \ref{fig:original_re}) and estimated (Figure \ref{fig:estim_re}) random intercepts in one of the runs of DGP 1 and shows how OMERF succesfully manages to capture the heterogeneoty at the group level. \\
Regarding the fixed effects, Figure \ref{fig:fix_cov_DGP1} shows the net association between the covariates and the response, by means of the partial plots extracted from the RF, giving an insight into the underlying latent process behind the ordinal model. It can be observed that the algorithm captures the quadratic and inverse linear trend in the variables \(x_1\) and \(x_2\), respectively.
%\textcolor{red}{2 note: bisogna definire qui e nella figura cosa intendiamo con response (asse y figura, la target è eta? se sì, specifichiamo. Togliere anche ALE dal grafico o definire cos'è). Poi, è ok che x4 e x5  abbiano praticamente effetto nullo? come x7?}

\begin{figure}[]
    \centering
    \subfigure[Distribution of random intercepts sampled from \( \mathcal{N}(0, 1)\).\label{fig:original_re}]{
        \includegraphics[scale=0.38]{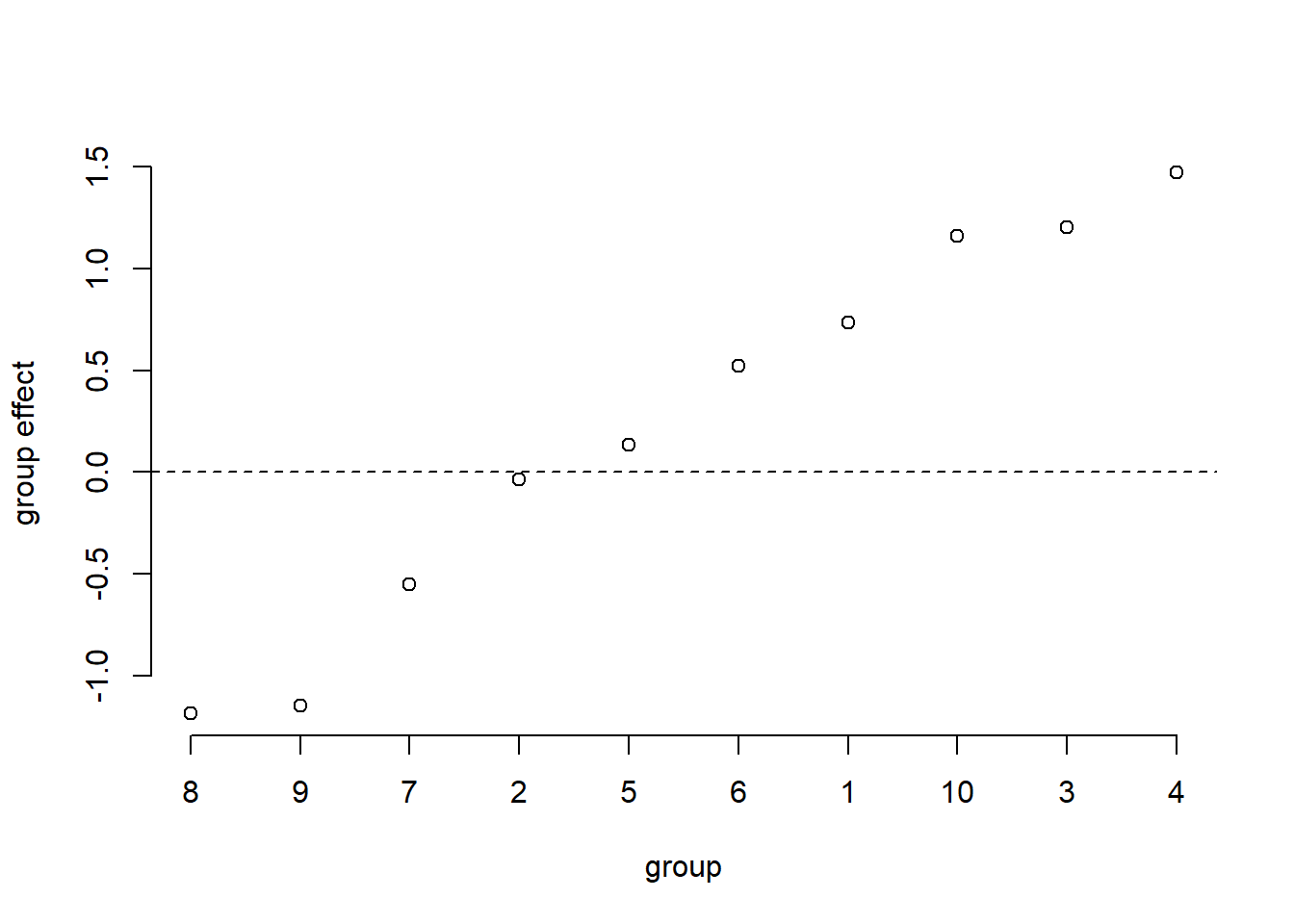}
    }
    \quad
    \subfigure[Distribution of random intercepts with their $95\%$ confidence intervals estimated by OMERF.\label{fig:estim_re}]{
        \includegraphics[scale=0.38]{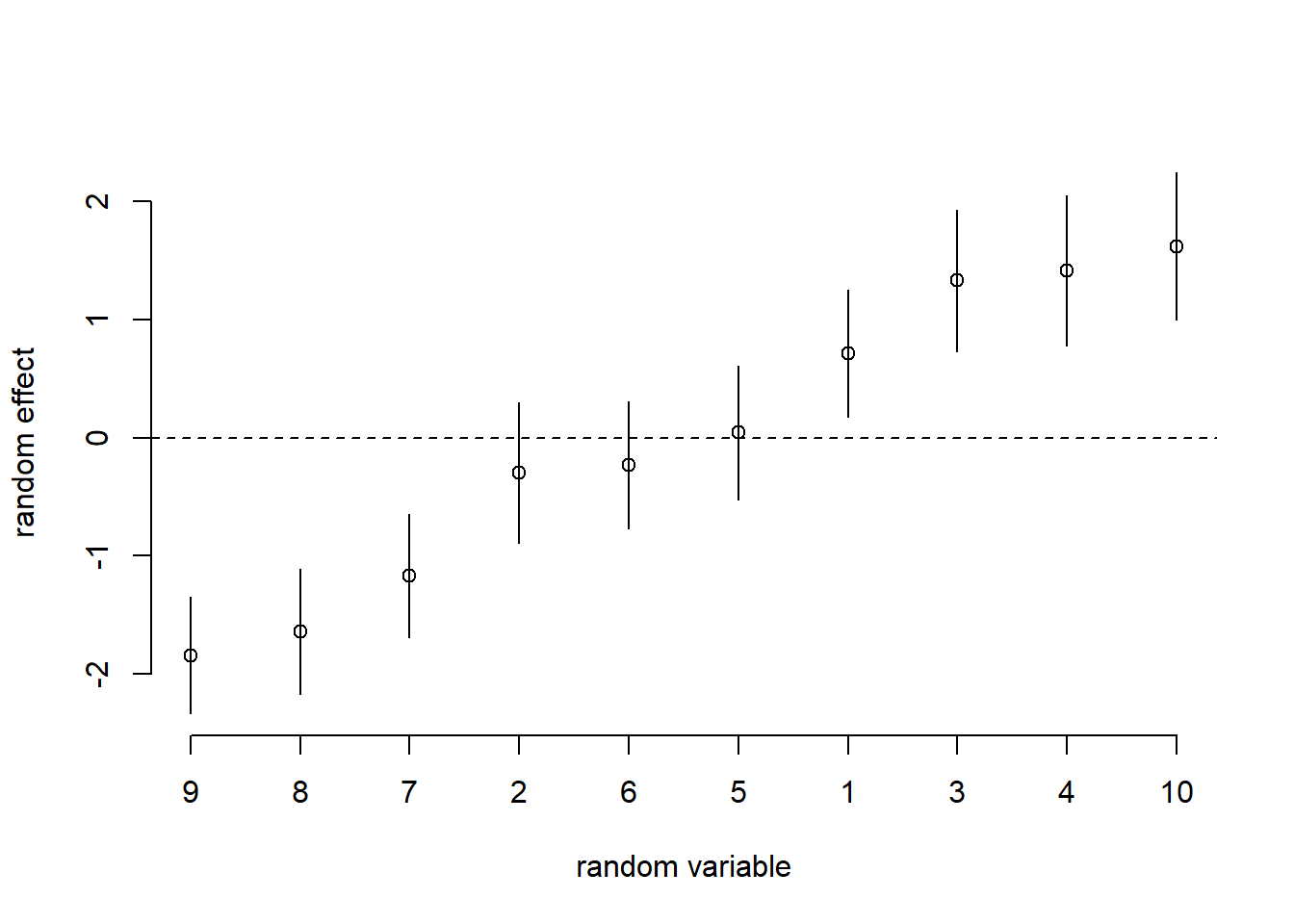}
    }
    \caption[]{Sampled and estimated random intercepts in one of the runs of DGP 1, described in Table \ref{table:DGPs}.}
    \label{fig:ranefDGP1}
\end{figure}

\begin{figure}[]
    \centering
    \includegraphics[width=0.8\textwidth]{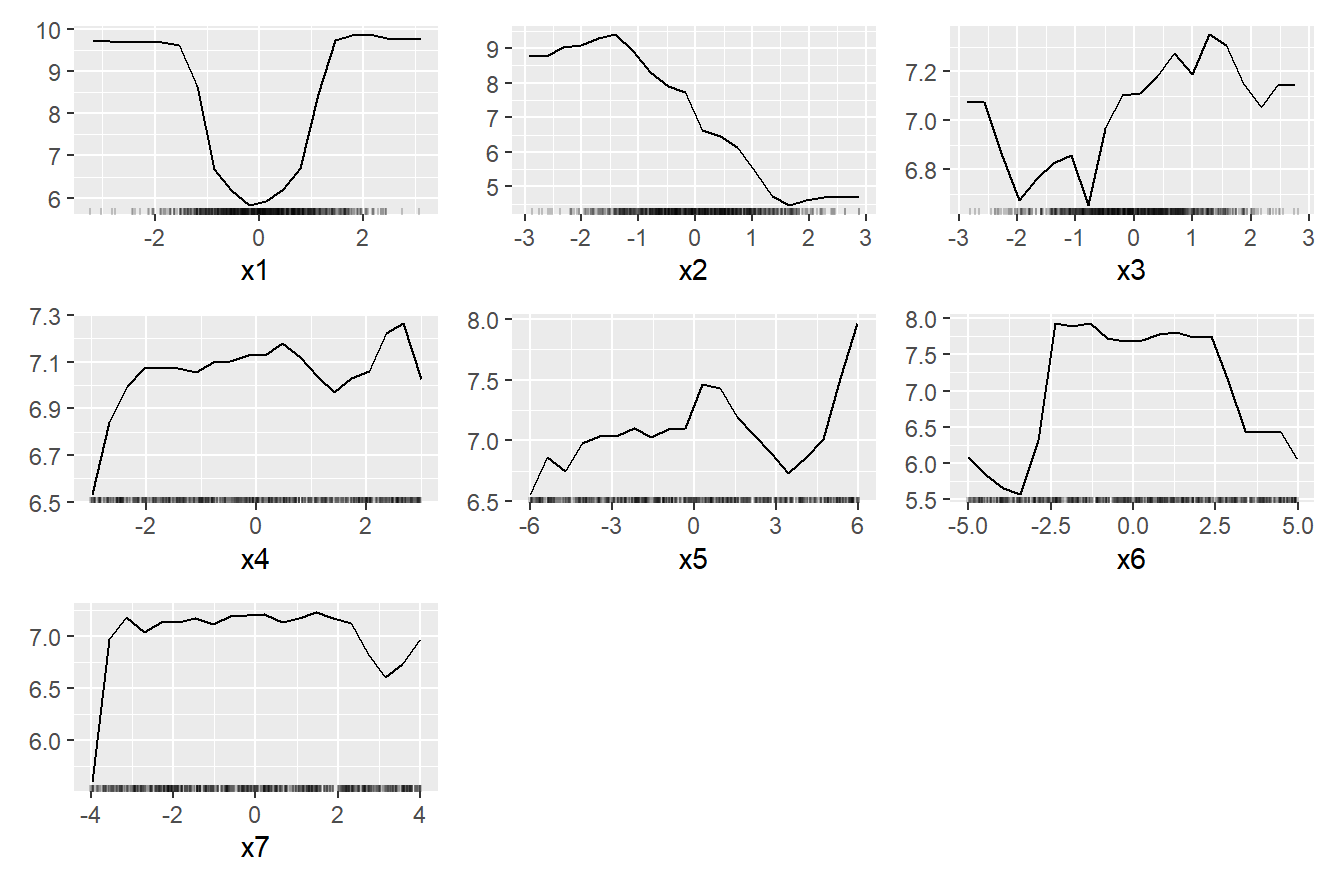}
    \caption{Partial plots for the fixed component of OMERF, for the seven covariates of DGP 1, described in Table \ref{table:DGPs}. The \(y\)-axis reports the increment/decrement of the target variable of the random forest in the iterative procedure, given the covariate on the \(x\)-axis.}
    \label{fig:fix_cov_DGP1}
\end{figure}

% Figure \ref{fig:ranefDGP1} recreates the differences in the hierarchical level (Figure \ref{fig:estim_re}) compared to the original random coefficients taken from the normal distribution (Figure \ref{fig:original_re}).
%Figure \ref{fig:fix_cov_DGP1} shows, on the other hand, the association of covariates and response, giving an insight into the underlying latent process behind the ordinal model. It can be observed that the algorithm captures the quadratic and inverse linear trend in the variables \(x_1\) and \(x_2\) respectively.}

\begin{longtable}[H]
    {|c c c c c c c c |}
    
    \hline
    
    % \rowcolor{blue!40}
    \textbf{DGP}& \textbf{Model} & \textbf{Acc} & \textbf{MSE} & \textbf{ARI} & \textbf{Cohen's k} & \textbf{Cardoso} & \textbf{Ballante}  \\
    \hline \hline\endfirsthead

    %\hline & \textbf{Model} & \textbf{Acc} & \textbf{MSE} & \textbf{ARI} & \textbf{Cohen's k} & \textbf{Cardoso} & \textbf{Ballante}  \\
    %\hline \hline
    \endhead

   \endfoot
 
\endlastfoot
    
    \textbf{1} & clm & 0.5729 & 0.7074 & 0.1228 & 0.1650 &  0.5498 & 0.2731  \\
    \textbf{} &  & (0.0019) & (0.0109) & (0.0018) & (0.0032) & (0.0022) & (0.0013)  \\
    \textbf{1} & clmm & 0.5713 & 0.7127 & 0.1176 & 0.1535 &  0.5503 & 0.2742  \\
    \textbf{} &  & (0.0018) & (0.0103) & (0.0017) & (0.0031) & (0.0021) & (0.0014)  \\
    \textbf{1} & ordforest & 0.6491 & 0.5076 & 0.2588 & 0.3346 & 0.4620 & 0.1991  \\
    \textbf{} &  & (0.0009) & (0.0046) & (0.0024) & (0.0039) & (0.0013) & (0.0007)  \\
    \textbf{1} & omerf & $\textcolor{blue}{0.6559}$ & $\textcolor{blue}{0.4506}$ & $\textcolor{blue}{0.2977}$ & $\textcolor{blue}{0.3963}$ & $\textcolor{blue}{0.4619}$ & $\textcolor{blue}{0.1983}$  \\
    \textbf{} &  & (0.0011) & (0.0037) & (0.0023) & (0.0028) & (0.0014) & (0.0006)  \\
    \hline
    \textbf{2} & clm & 0.7496 & 0.5047 & 0.1923 & 0.2042 & 0.3811 & 0.1704  \\
    \textbf{} &  & (0.0008) & (0.0067) & (0.0035) & (0.0036) & (0.0016) & (0.0032)  \\
    \textbf{2} & clmm & 0.7512 & 0.5061 & 0.1812 & 0.1962 & 0.3791 & 0.1806  \\
    \textbf{} &  & (0.0008) & (0.0063) & (0.0033) & (0.0035) & (0.0015) & (0.0040)  \\
    \textbf{2} & ordforest & 0.8025 & 0.3148 & 0.3997 & 0.3954 & 0.2909 & $\textcolor{blue}{0.0843}$  \\
    \textbf{} &  & (0.0006) & (0.0026) & (0.0042) & (0.0047) & (0.0011) & (0.0003)  \\
    \textbf{2} & omerf & $\textcolor{blue}{0.8089}$ & $\textcolor{blue}{0.2565}$ & $\textcolor{blue}{0.5330}$ & $\textcolor{blue}{0.5066}$ & $\textcolor{blue}{0.2836}$ & 0.0899  \\
    \textbf{} &  & (0.0007) & (0.0021) & (0.0036) & (0.0037) & (0.0013) & (0.0004)  \\
    \hline
    \textbf{3} & clm & 0.6238 & 0.6589 & 0.2182 & 0.2943 & 0.5082 & 0.2314  \\
    \textbf{} &  & (0.0031) & (0.0197) & (0.0043) & (0.0054) & (0.0041) & (0.0025)  \\
    \textbf{3} & clmm & 0.6236 & 0.6646 & 0.2156 & 0.2893 & 0.5082 & 0.2330  \\
    \textbf{} &  & (0.0031) & (0.0202) & (0.0044) & (0.0059) & (0.0041) & (0.0025)  \\
    \textbf{3} & ordforest & 0.6705 & 0.5371 & 0.2853 & 0.3576 & 0.4490 & 0.1829  \\
    \textbf{} &  & (0.0016) & (0.0079) & (0.0038) & (0.0082) & (0.0022) & (0.0010)  \\
    \textbf{3} & omerf & $\textcolor{blue}{0.6748}$ & $\textcolor{blue}{0.4589}$ & $\textcolor{blue}{0.3334}$ & $\textcolor{blue}{0.4237}$ & $\textcolor{blue}{0.4464}$ & $\textcolor{blue}{0.1792}$  \\
    \textbf{} &  & (0.0016) & (0.0050) & (0.0032) & (0.0036) & (0.0020) & (0.0008)  \\    
    \hline
    \textbf{4} & clm & 0.7532 & 0.5372 & 0.2389 & 0.2725 & 0.3843 & 0.2100  \\
    \textbf{} &  & (0.0015) & (0.0143) & (0.0036) & (0.0043) & (0.0031) & (0.0069)  \\
    \textbf{4} & clmm & 0.7536 & 0.5377 & 0.2327 & 0.2659 & 0.3839 & 0.2118  \\
    \textbf{} &  & (0.0015) & (0.0138) & (0.0036) & (0.0047) & (0.0030) & (0.0071)  \\
    \textbf{4} & ordforest & 0.7953 & 0.3708 & 0.3583 & 0.3722 & 0.3102 & 0.0998  \\
    \textbf{} &  & (0.0009) & (0.0059) & (0.0075) & (0.0080) & (0.0019) & (0.0010)  \\
    \textbf{4} & omerf & $\textcolor{blue}{0.8069}$ & $\textcolor{blue}{0.2865}$ & $\textcolor{blue}{0.5126}$ & $\textcolor{blue}{0.5063}$ & $\textcolor{blue}{0.2914}$ & $\textcolor{blue}{0.0869}$  \\
    \textbf{} &  & (0.0011) & (0.0039) & (0.0043) & (0.0036) & (0.0022) & (0.0005)  \\
    \hline
    \textbf{5} & clm & 0.5598 & 0.7261 & 0.0970 & 0.1288 & 0.5604 & 0.2819  \\
    \textbf{} &  & (0.0009) & (0.0045) & (0.0012)  & (0.0026)  & (0.0008)  & (0.0006)  \\
    \textbf{5} & clmm & 0.5586 & 0.7315 & 0.0940 & 0.1206 & 0.5611 & 0.2851  \\
    \textbf{} &  & (0.0009) & (0.0048) & (0.0014)  & (0.0029)  & (0.0021)  & (0.0009)  \\
    \textbf{5} & ordforest & $\textcolor{blue}{0.6409}$ & 0.5002 & 0.2541 & 0.3283 & $\textcolor{blue}{0.4659}$ & $\textcolor{blue}{0.2038}$  \\
    \textbf{} &  & (0.0009) & (0.0033) & (0.0025)  & (0.0033)  & (0.0013)  & (0.0005)  \\
    \textbf{5} & omerf & 0.6342 & $\textcolor{blue}{0.4938}$ & $\textcolor{blue}{0.2645}$ & $\textcolor{blue}{0.3623}$ & 0.4866 & 0.2166  \\
    \textbf{} &  & (0.0012) & (0.0041) & (0.0033)  & (0.0032)  & (0.0014)  & (0.0005)  \\
    \hline
    \textbf{6} & clm & 0.7489 & 0.4900 & 0.1803 & 0.1854 & 0.3786 & 0.1653  \\
    \textbf{} &  & (0.0005) & (0.0036) & (0.0038)  & (0.0039)  & (0.0008)  & (0.0025)  \\
    \textbf{6} & clmm & 0.7503 & 0.4927 & 0.1699 & 0.1812 & 0.3776 & 0.1770  \\
    \textbf{} &  & (0.0005) & (0.0041) & (0.0038)  & (0.0038)  & (0.0008)  & (0.0034)  \\
    \textbf{6} & ordforest & 0.8063 & 0.2986 & 0.4187 & 0.4089 & 0.2834 & $\textcolor{blue}{0.0824}$  \\
    \textbf{} &  & (0.0004) & (0.0023) & (0.0040)  & (0.0049)  & (0.0008)  & (0.0003)  \\
    \textbf{6} & omerf & $\textcolor{blue}{0.8104}$ & $\textcolor{blue}{0.2569}$ & $\textcolor{blue}{0.5395}$ & $\textcolor{blue}{0.5119}$ & $\textcolor{blue}{0.2830}$ & 0.0910  \\
    \textbf{} &  & (0.0007) & (0.0021) & (0.0042)  & (0.0043)  & (0.0013)  & (0.0003)  \\
    \hline
    \textbf{7} & clm & 0.5696 & 0.7339 & 0.1205 & 0.1691 & 0.5571 & 0.2791  \\
    \textbf{} &  & (0.0015) & (0.0133) & (0.0020)  & (0.0039)  & (0.0019)  & (0.0020)  \\
    \textbf{7} & clmm & 0.5643 & 0.7491 & 0.1119 & 0.1571 & 0.5630 & 0.2848  \\
    \textbf{} &  & (0.0018) & (0.0127) & (0.0021)  & (0.0045)  & (0.0021)  & (0.0022)  \\
    \textbf{7} & ordforest & $\textcolor{blue}{0.6414}$ & 0.5334 & 0.2554 & 0.3254 & $\textcolor{blue}{0.4719}$ & $\textcolor{blue}{0.2074}$  \\
    \textbf{} &  & (0.0011) & (0.0047) & (0.0022)  & (0.0031)  & (0.0015)  & (0.0008)  \\
    \textbf{7} & omerf & 0.6339 & $\textcolor{blue}{0.5121}$ & $\textcolor{blue}{0.2714}$ & $\textcolor{blue}{0.3589}$ & 0.4882& 0.2169  \\
    \textbf{} &  & (0.0016) & (0.0071) & (0.0035)  & (0.0036)  & (0.0021)  & (0.0012)  \\
    \hline
    \textbf{8} & clm & 0.7507 & 0.4943 & 0.2032 & 0.2139 & 0.3774 & 0.1729  \\
    \textbf{} &  & (0.0008) & (0.0073) & (0.0040)  & (0.0040)  & (0.0017)  & (0.0036)  \\
    \textbf{8} & clmm & 0.4985 & 0.4927 & 0.1908 & 0.2029 & 0.3789 & 0.1809  \\
    \textbf{} &  & (0.0008) & (0.0069) & (0.0042)  & (0.0047)  & (0.0016)  & (0.0040)  \\
    \textbf{8} & ordforest & 0.8011 & 0.3235 & 0.4025 & 0.3995 & $\textcolor{blue}{0.2947}$ & $\textcolor{blue}{0.0885}$  \\
    \textbf{} &  & (0.0005) & (0.0032) & (0.0047)  & (0.0041)  & (0.0011)  & (0.0003)  \\
    \textbf{8} & omerf & $\textcolor{blue}{0.8018}$ & $\textcolor{blue}{0.2766}$ &  $\textcolor{blue}{0.5195}$ & $\textcolor{blue}{0.4942}$ & 0.2955 & 0.0965  \\
    \textbf{} &  & (0.0007) & (0.0034) & (0.0032)  & (0.0031)  & (0.0016)  & (0.0004)  \\
    \hline
    \textbf{9} & clm & 0.8711 & 0.1610 & 0.7291 & 0.7761 & 0.1979 & 0.0419  \\
    \textbf{} &  & (0.0005) & (0.0016) & (0.0021)  & (0.0015)  & (0.0012)  & (0.0001)  \\
    \textbf{9} & clmm & $\textcolor{blue}{0.8716}$ & $\textcolor{blue}{0.1609}$ & $\textcolor{blue}{0.7297}$ & $\textcolor{blue}{0.7769}$ & $\textcolor{blue}{0.1977}$ & $\textcolor{blue}{0.0417}$  \\
    \textbf{} &  & (0.0004) & (0.0014) & (0.0019)  & (0.0012)  & (0.0009)  & (0.0001)  \\
    \textbf{9} & ordforest & 0.8544 & 0.2145 & 0.6858 & 0.7414 & 0.2262 & 0.0484  \\
    \textbf{} &  & (0.0004) & (0.0024) & (0.0022)  & (0.0013)  & (0.0012)  & (0.0002)  \\
    \textbf{9} & omerf & 0.8389 & 0.2704 & 0.7127& 0.6167 & 0.2510 & 0.0818  \\
    \textbf{} &  & (0.0004) & (0.0028) & (0.0024)  & (0.0011)  & (0.0011)  & (0.0002)  \\
    \hline
    \textbf{10} & clm & $\textcolor{blue}{0.8756}$ & $\textcolor{blue}{0.1559}$ & $\textcolor{blue}{0.7376}$ & $\textcolor{blue}{0.7824}$ & $\textcolor{blue}{0.1919}$ & 0.0386  \\
    \textbf{} &  & (0.0004) & (0.0012) & (0.0019)  & (0.0013)  & (0.0009)  & (9.1742e-05)  \\
    \textbf{10} & clmm & 0.8755 & $\textcolor{blue}{0.1559}$ & $\textcolor{blue}{0.7376}$ & 0.7821 & 0.1920 & $\textcolor{blue}{0.0385}$  \\
    \textbf{} &  & (0.0004) & (0.0011) & (0.0017)  & (0.0012)  & (0.0009)  & (9.0919e-05)  \\
    \textbf{10} & ordforest & 0.8420 & 0.2610 & 0.6492 & 0.7154 & 0.2486 & 0.0523  \\
    \textbf{} &  & (0.0005) & (0.0031) & (0.0025)  & (0.0013)  & (0.0012)  & (0.0001)  \\
    \textbf{10} & omerf & 0.8214 & 0.3340 & 0.6004 & 0.6779 & 0.2845 & 0.0911  \\
    \textbf{} &  & (0.0005) & (0.0054) & (0.0037)  & (0.0015)  & (0.0015)  & (0.0005)  \\
    \hline

    \caption{Mean and variances of prediction performances, measured by six indices, of the four compared methods across 100 runs of the 10 DGPs listed in Table \ref{table:DGPs}.}
    \label{table:res_lin}
\end{longtable}

\section{Case study}
\label{sec:case}
In this section, we delve into a real-world application of the OMERF method in learning analytics. %, conducting an analysis with the objective of comparing its performance with the models delineated in Section \ref{sec:methods}.

\subsection{The dataset}
\label{sec:dataset}

The data employed in the study concern 15-year-old students attending Italian schools who completed the PISA 2022 survey questionnaire.
% \CM{aggiungere reference a pisa}
%(\cite{organisation2023pisa}). 
The dataset comprises information about 10,552 students across 340 schools. After filtering for complete cases and schools with more than 10 students, the processed dataset consists of 7,639 observations, representing students enrolled in 293 schools. For model training and evaluation, a random sample comprising 80$\%$ of the observations is designated as the training set, while the remaining 20$\%$ constitutes the test set.

The objective of this case study is to evaluate the performance of OMERF in predicting and modelling students' mathematical performance, accounting for student characteristics and attended schools. In accordance with the threshold established by the OECD, the output variable, that is the student PISA score, is categorized into three ordinal levels: the lowest level encompasses students classified in levels 1 or 2, while the highest level includes those achieving levels 5 or 6, with the remaining students falling within the intermediate class \footnote{Levels 5 and 6 are designated to students with high abilities, levels 3 and 4 to students with moderate abilities, and the remaining two levels to those with basic or no abilities. For more details about the PISA proficiency levels, please refer to \texttt{https://www.oecd.org/pisa/.}}.
%\CM{Possiamo magari aggiungere in una footnote qualche spiegazione di questi livelli, tipo che 1 e 2 sono ritenuti innumeracy?}. 
% \textcolor{red}{qui di seguito alleggerirei il testo dicendo solo che le variabili a livello studente utilizzate per prevedere i math score, che riguardano demographic, family background ecc ecc, sono listate nella tabella. Aggiungerei direttamente nella tabella una micro spiegazione in più. per le categoriche nella tabella riporterei solo le percentuali (non le freq assolute). per gender e per tutte le cat dire cosa è 1 e 0.}
The student-level variables, extracted from the OECD-PISA database, used to predict the mathematics test scores include demographic factors, educational indicators, family background information, factors related to home and school environment, and self-perception attributes.
% Demographic factors, such as gender and immigrant status, alongside educational indicators, like grade level and family background - including parents' highest education level and ESCS - are considered as covariates for prediction. Additionally, factors related to home environment, such as access to and quality of internet services and video games, as well as the number of hours spent on homework per week, and perceived support from both teachers and parents, are incorporated into the analysis. School climate aspects, including the perception of school risk, sense of belonging, and experiences of bullying, are also used. Furthermore, self-perception attributes like cooperation, perseverance, assertiveness, empathy, emotional control, stress resistance, and curiosity are included as relevant predictor variables.
A detailed description of these variables, along with corresponding descriptive statistics, is provided in Table \ref{table:var_stud}. All indicators variables are build by PISA by combining multiple responses to questionnaires in numerical indicators.

\footnotesize
\begin{longtable}[]{|p{8.5em}|p{12em}|p{7em}|p{9em}|}

    \hline
    \textbf{Variable name} & \textbf{Variable description} & \textbf{Variable type} & \textbf{Distribution} \\
    \hline \hline \endfirsthead

    \hline
    \textbf{Variable name} & \textbf{Variable description} & \textbf{Variable type} & \textbf{Distribution}   \\
    \hline \hline \endhead

    \hline \multicolumn{4}{|r|}{{Continued on next page}} \\ \hline
\endfoot

\endlastfoot
    %\(stud_ID\) &  Student identification number &  factor & 7639 unique values & & & &  \\
   % \hline
   % \(school_ID\) &  School identification number &  factor & 393 levels  & & & &  \\
   % \hline
    mate3 &  Output variable, mathematics test score & ordered factor & 1:48$\%$\hspace{1.8cm} 2:44$\%$\hspace{1.8cm} 3:8$\%$\\
    \hline
    gender &  Student gender (0 = male, 1 = female)& factor & 1:53$\%$\hspace{1.8cm} 0:47$\%$   \\
    \hline
    immig &  Student immigration status (0 = native Italian; 1 = $1^{st}$-generation immigrant; 2 = $2^{nd}$-generation immigrant) &  factor & 0:89$\%$\hspace{1.8cm} 1:8$\%$\hspace{1.8cm}2:3$\%$   \\
    \hline
    grade &  School grade attended (10 = regular student; 9 = late enrolled student; 11 = early enrolled student)&  factor & 9:10$\%$\hspace{1.8cm} 10:85$\%$\hspace{1.8cm} 11:5$\%$ \\
    \hline
    video\_games &  Indicator of frequency of use at home of video or online games &  numeric & Mean:3.27; SD:1.59; \hspace{0.4cm}Range: [1.00;6.00]  \\
    \hline
    internet\_quality &  Indicator of quality of access to ICT at school &  numeric & Mean:-0.18; SD:0.83; \hspace{0.3cm}Range: [-2.80;2.89]  \\
    \hline
    internet\_availability &  Indicator of ICT availability outside of school & numeric & Mean:5.70; SD:0.91; Range: [0.00;6.00]  \\
    \hline
    SCHRISK &  Indicator of perceived school safety risks &  numeric & Mean:0.01; SD:0.82; \hspace{0.3cm}Range: [-0.46;3.05]  \\
    \hline
    BULLIED&  Indicator of being bullied &  numeric & Mean:-0.47; SD:0.87; \hspace{0.3cm}Range: [-1.23;4.69]  \\
    \hline
    BELONG &  Indicator of sense of belonging to school & numeric & Mean:0.01; SD:0.89; \hspace{0.3cm}Range: [-3.26;2.78]   \\
    \hline
    COOPAGR &   Indicator of cooperation (agreement) &  numeric & Mean:0.12; SD:1.00;\hspace{0.3cm} Range: [-5.24;6.13]   \\
    \hline
    TEACHSUP &   Indicator of perceived teachers support &  numeric & Mean:-0.20; SD:1.11; \hspace{0.3cm}Range: [-2.91;1.56]  \\
    \hline
    FAMSUP &   Indicator of perceived family support &  numeric & Mean:-0.02; SD:0.93; \hspace{0.3cm}Range: [-3.01;1.96]   \\
    \hline
    PERSEVAGR &   Indicator of perseverance (agreement) &  numeric & Mean:0.07; SD:0.98; \hspace{0.3cm}Range: [-5.91;4.89]  \\
    \hline
    ASSERAGR &   Indicator of assertiveness (agreement) &  numeric & Mean:-0.03; SD:1.01; \hspace{0.3cm}Range: [-8.23;7.23]  \\
    \hline
    EMPATAGR &   Indicator of empathy (agreement) &  numeric & Mean:0.01; SD:0.99; \hspace{0.3cm}Range: [-6.46;4.69]   \\
    \hline
    EMOCOAGR &   Indicator of emotional control (agreement) &  numeric & Mean:-0.09; SD:0.98; \hspace{0.3cm}Range: [-5.17;5.58]  \\
    \hline
    STRESAGR &  Indicator of stress resistance (agreement) &  numeric & Mean:-0.18; SD:1.00; \hspace{0.3cm}Range: [-5.26;5.49]   \\
    \hline
    CURIOAGR &   Indicator of curiosity (agreement) &  numeric & Mean:0.09; SD:0.96; \hspace{0.3cm}Range: [-4.95;4.18]   \\
    \hline
    study\_time &  Total time for all homework
in all subjects per week &  numeric & Mean:3.50; SD:1.51; \hspace{0.3cm}Range: [1.00;6.00]   \\
    \hline
    HISCED &  Highest level of education of parents &  numeric & Mean:7.03; SD:2.14;\hspace{0.3cm} Range: [1.00;10.00]   \\
    \hline
    ESCS &  socio-economic family index &  numeric & Mean:-0.01; SD:0.87;\hspace{0.3cm} Range: [-3.23;2.78] \\
    \hline
    \caption{Student-level variables extracted from the OECD-PISA database. }
    \label{table:var_stud}
\end{longtable}

\normalsize
\subsection{Model results}
\label{sec:modres}

% \textcolor{red}{manca qui una frase per dire che modello costruiamo: usiamo tutte le tot var a livello studente listate in tabella come fixed effects covariates, inseriamo una intercetta random per account for the school effect e che parametri di OMERF vengono settati (convergenza, number of trees. ecc) e così per gli altri metodi che confrontiamo. In tabella 4, colorerei in colore diverso il secondo classificato, che dovrebbe essere omerf a occhio}

We run OMERF to predict the student mathematics level \texttt{mate3}, that is the ordinal target variable, by including all other variables listed in Table \ref{table:var_stud} as fixed-effects covariates and considering a random intercept ($Q=0$) to estimate the school-effect \citep{raudenbush1995estimation}. We compare the performance of OMERF with the ones of CLM, CLMM and Ordinal Forest\footnote{OMERF and the ordinal forest are run with default inputs.}. For CLM and Ordinal Forest, the random intercept is not considered.

In Table \ref{table:res_math}, the predictive performance (computed on the test set) of the four methods are reported.

\begin{table}[H]
    \centering 
    \begin{tabular}{| c c c c c c c |}
    \hline
    % \rowcolor{blue!40}
     \textbf{Model} & \textbf{Acc} & \textbf{MSE} & \textbf{ARI} & \textbf{Cohen's k} & \textbf{Cardoso} & \textbf{Ballante}  \\
    \hline \hline
     clm & 0.5938 & 0.4573 & 0.0739 & 0.2484 & 0.5073 & 0.2835  \\
      clmm & $\textcolor{blue}{0.6699}$  & $\textcolor{blue}{0.3339}$ &  $\textcolor{blue}{0.1884}$ & $\textcolor{blue}{0.3963}$ & $\textcolor{blue}{0.4228}$ & $\textcolor{blue}{0.2121}$  \\
     ordforest & 0.6010 & 0.4383 & 0.0914 & 0.2597 & 0.4937 & 0.2607  \\
     omerf & \textcolor{teal}{0.6444} & \textcolor{teal}{0.3734} & \textcolor{teal}{0.1496} & \textcolor{teal}{0.3546} & \textcolor{teal}{0.4527} & \textcolor{teal}{0.2402}  \\
    \hline
    %\textbf{2} & clm & 0.6082 & 0.4272 & 0.0945 & 0.2764 & 0.4909 & 0.2678  \\
    %\textbf{2} & clmm & $\textcolor{blue}{0.6745}$  & $\textcolor{blue}{0.3333}$ & $\textcolor{blue}{0.1958}$ & $\textcolor{blue}{0.4061}$ & $\textcolor{blue}{0.4187}$ & $\textcolor{blue}{0.2191}$  \\
    %\textbf{2} & ordforest & 0.6358 & 0.4015 & 0.1367 & 0.3247 & 0.4586 & 0.2409  \\
   % \textbf{2} & omerf & 0.6450 & 0.3766 & 0.1528 & 0.3572 & 0.4549 & 0.2369  \\
   % \hline
    \end{tabular}
    \\[10pt]
    \caption{Prediction performances (computed on the test set) of the four compared methods applied to the real-world case study.}
    \label{table:res_math}
\end{table}

CLMM consistently outperforms other methods across all metrics considered, closely followed by OMERF, which achieves slightly lower but comparable results. %{it can be noticed that the CLMM and OMERF algorithms shows comparable performance indices and outperform other algorithms across all considered metrics.} %In particular [inserire qualche esempio]. 
This underscores the importance of accounting for the hierarchical structure of the data. The fact that CLMM performs slightly better than OMERF could be explained by the presence of a strong linear relationship between predictors and response. To check that, Figures \ref{fig:varimp} and \ref{fig:par_plot_case} illustrate the variable importance plot (VIMP) and the partial plots produced by OMERF,
% \CM{FORSE NEI METODI VALE LA PENA METTERE UNA FRASE IN CUI DICIAMO CHE LA STIMA DELLA FIXED COMPONENT DI FATTO E' UN RF OBJECT (SE GIà NON LO DICIAMO ESPLICITAMENTE) E QUINDI TUTTA LA PARTE DI INTERPRET (PARTIAL PLOTS E IMPOPLOT ECC) VIENE COME SEMPRE}
that enable examination of the relationship between predictors and the output variable. %{Moreover, the OMERF algorithm allows to interpret the results according to the random forest tools. Variable importance analysis, depicted in Figure \ref{fig:varimp}, reveals the most important variables in the prediction.} 
From the VIMP (\ref{fig:varimp}), we observe that ESCS is the most important predictor, consistent with existing literature. In the partial plot (\ref{fig:par_plot_case}), ESCS demonstrates a quasi-linear relationship with the response, potentially explaining the comparable performance of OMERF and CLMM, as linear models can adequately capture such relationships. %Additionally,  Some variables like [inserire alcune non lineari] present a non linear relationship that OMERF, in contrast with CLMM, is able to capture. 
On the other hand, Figure \ref{fig:par_plot_case} highlights the nonlinear association between some of the most important variables, such as \textit{EMOCOAGR}, \textit{CURIOAGR} and \textit{FAMSUP}, and the output variable. For instance, for what concerns the student emotional control (\textit{EMOCOAGR}), we observe a steep increment of the response when this covariate exceeds its average value of 0, while we observe slight variations in the response for both high and low values of the covariate, with high values being associated to higher values of the response. Concerning student curiosity (\textit{CURIOAGR}), we observe a clear increase in the response when the value of this covariate moves from 0 to the right limit of its range, while we observe no significant variations across its negative values. One of the strengths of OMERF is its ability to capture these types of relationships, which CLMM can not. However, the variable importance plot indicates that these covariates are not the most influential variables. This, combined with the quasi-linear relationship between ESCS and the output variable, likely explains the higher performance of CLMM compared to OMERF.

%\begin{figure}[H]
%    \centering
%    \subfigure[caption.]{
%        \includegraphics[scale=0.3]%{Images/varimp.png}  }    \quad
%    \subfigure[ESCS.]{
%        \includegraphics[scale=0.3]{Images/varimp_ESCS.png}
%    }
%    \quad
%   \label{fig:varimp}
%\end{figure}

\begin{figure}[H]
    \centering
    \includegraphics[width=0.9\textwidth]{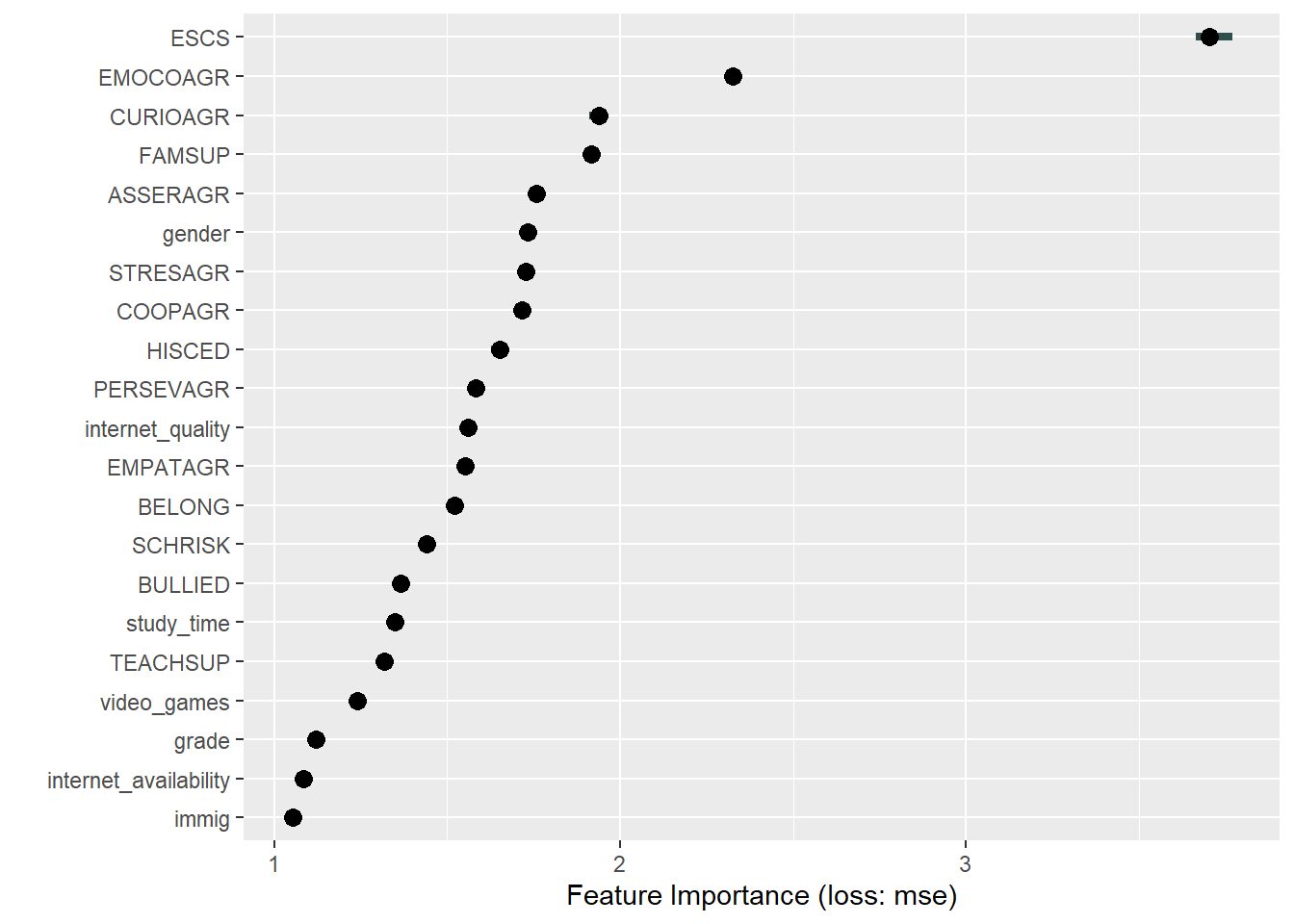}
    \caption{Variable Importance plot for the fixed component of OMERF, in the real-world case study.}
    \label{fig:varimp}
\end{figure}

%\begin{figure}[H]
%    \centering
%    \includegraphics[width=0.8\textwidth]{Images/parplot.png}
%    \caption{}
%    \label{fig:fix_cov_DGP1}
%\end{figure}

\begin{figure}
\fontsize{6}{8}\selectfont
\centering
    \subfigure[ESCS]{\resizebox*{4.1cm}{!}{\includegraphics{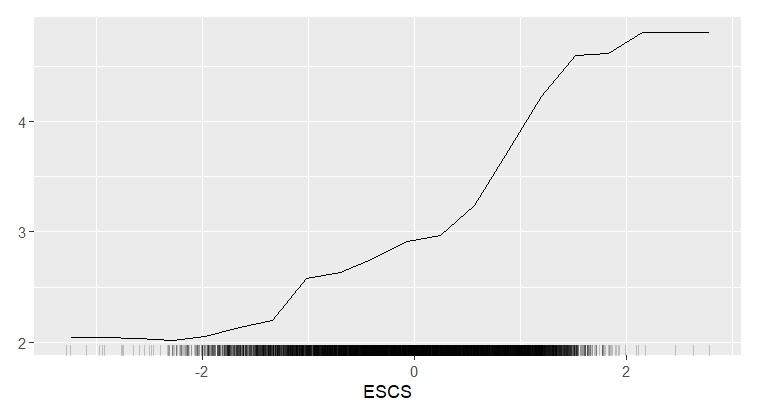}}}
    \hspace{0.3cm}
    \subfigure[EMOCOAGR]{\resizebox*{4.1cm}{!}{\includegraphics{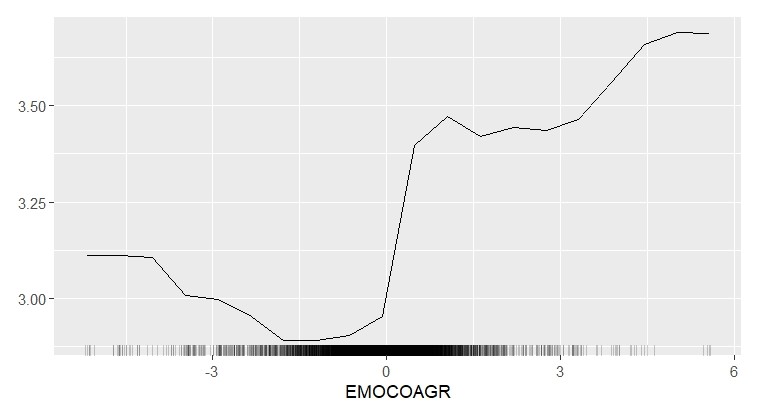}}}
    \hspace{0.3cm}
    \subfigure[CURIOAGR]{\resizebox*{4.1cm}{!}{\includegraphics{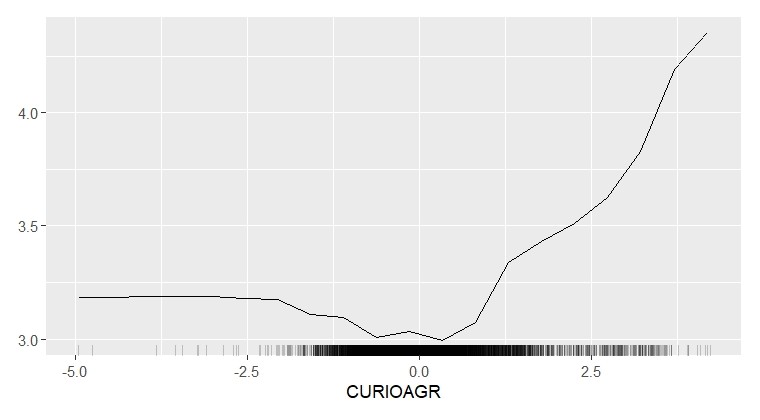}}}
    \hspace{0.3cm}
     \subfigure[FAMSUP]{\resizebox*{4.1cm}{!}{\includegraphics{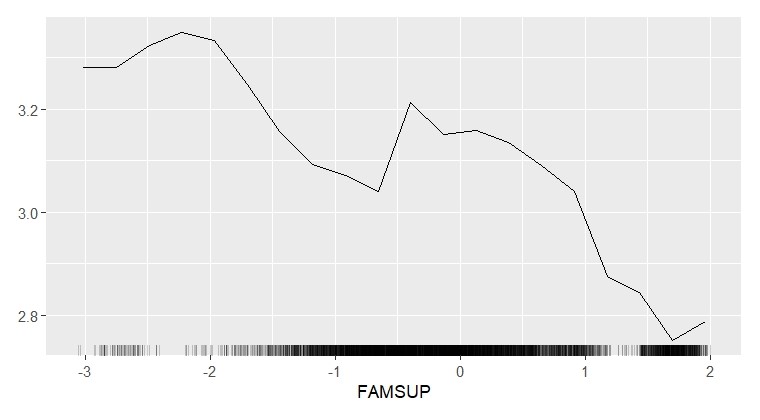}}}
    \hspace{0.3cm}
    \subfigure[ASSERAGR]{\resizebox*{4.1cm}{!}{\includegraphics{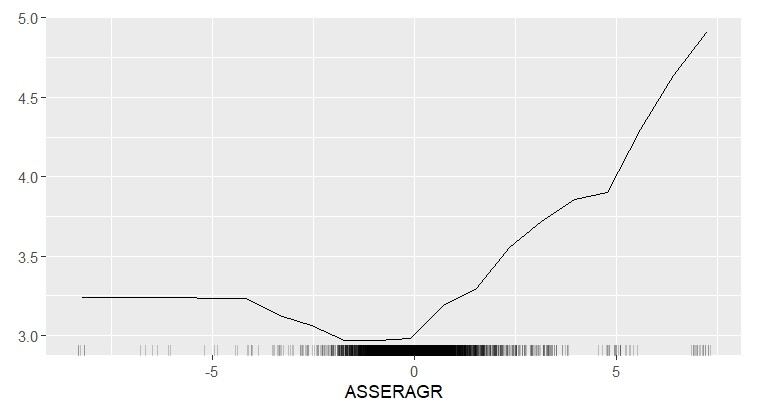}}}
    \hspace{0.3cm}
    \subfigure[gender]{\resizebox*{4.1cm}{!}{\includegraphics{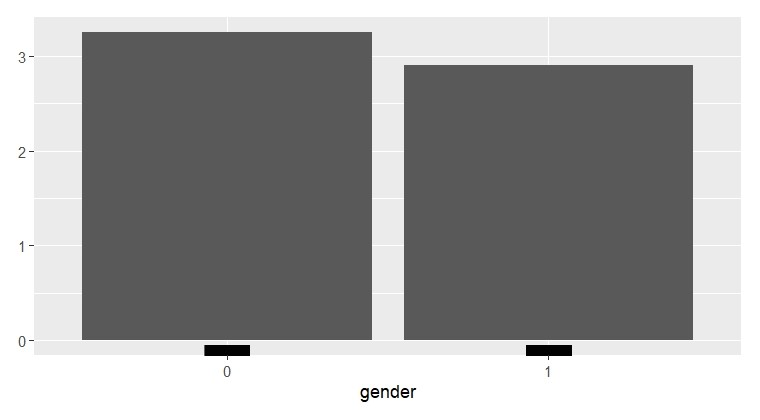}}}
    \hspace{0.3cm}
    \subfigure[STRESAGR]{\resizebox*{4.1cm}{!}{\includegraphics{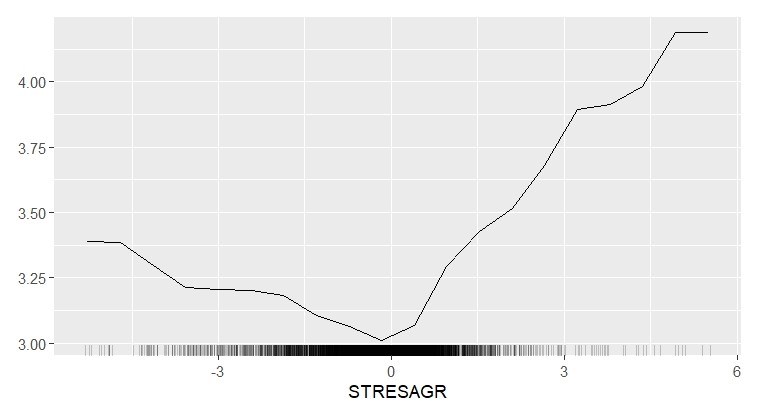}}}
    \hspace{0.3cm}
    \subfigure[COOPAGR]{\resizebox*{4.1cm}{!}{\includegraphics{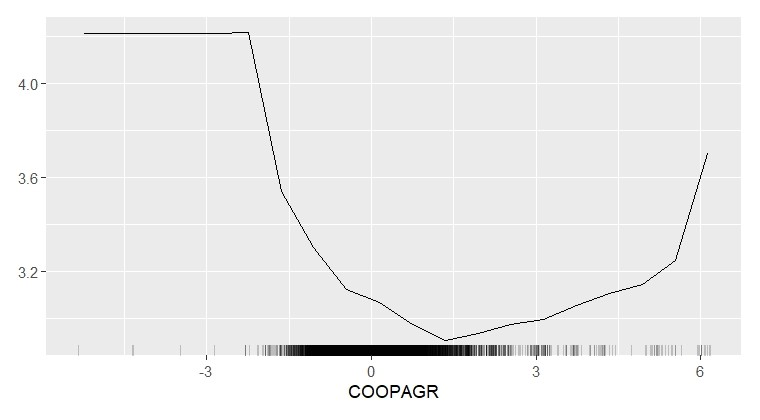}}}
    \hspace{0.3cm}
    \subfigure[HISCED]{\resizebox*{4.1cm}{!}{\includegraphics{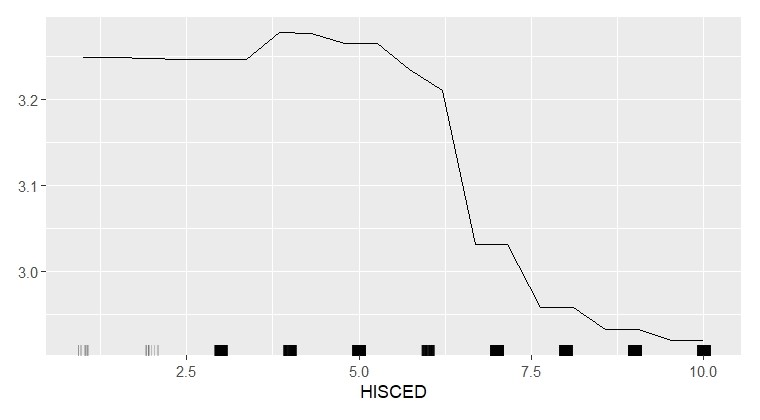}}}
    \hspace{0.3cm}
    \subfigure[PERSEVAGR]{\resizebox*{4.1cm}{!}{\includegraphics{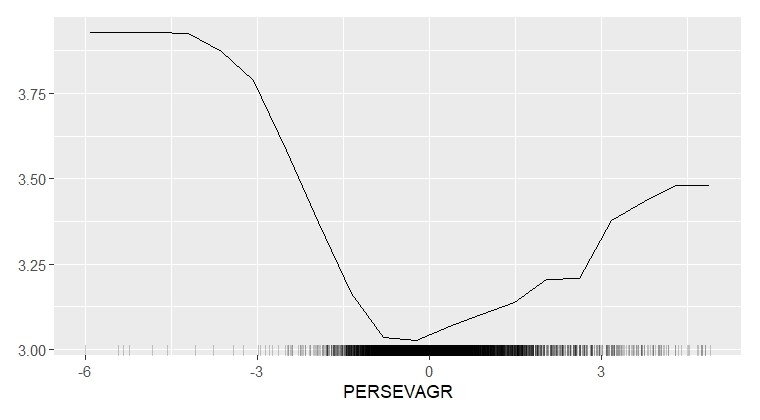}}}
    \hspace{0.3cm}
    \subfigure[internet\_quality]{\resizebox*{4.1cm}{!}{\includegraphics{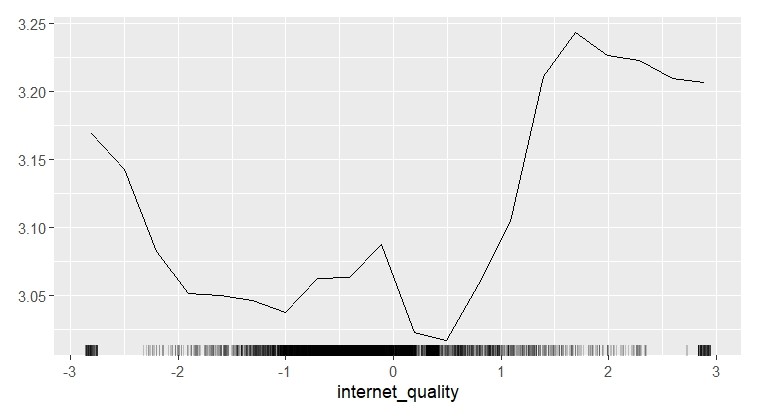}}}
    \hspace{0.3cm}
    \subfigure[EMPATAGR]{\resizebox*{4.1cm}{!}{\includegraphics{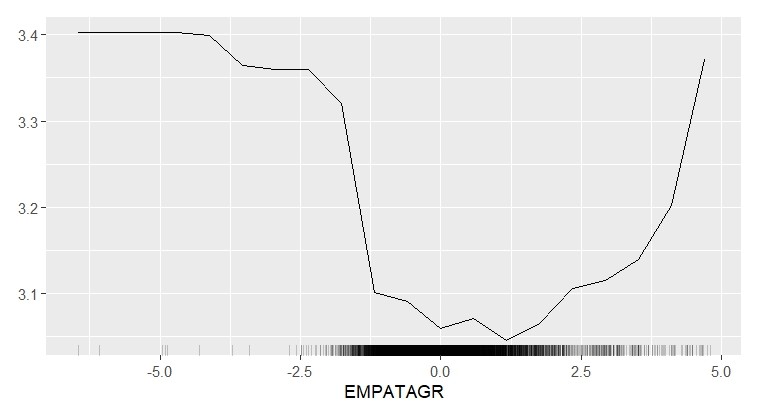}}}
    \hspace{0.3cm}
    \subfigure[BELONG]{\resizebox*{4.1cm}{!}{\includegraphics{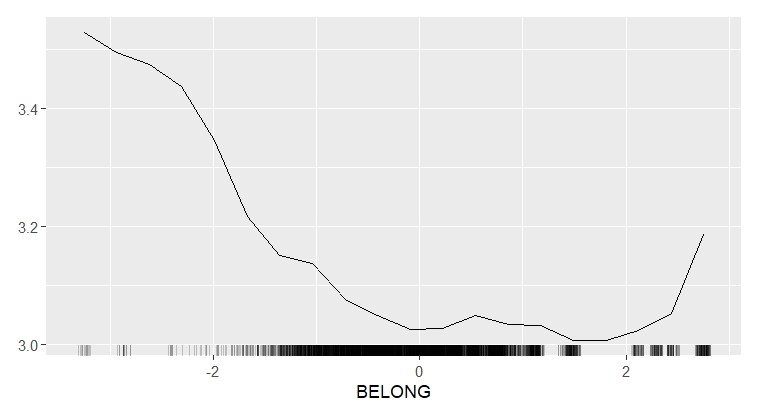}}}
    \hspace{0.3cm}
    \subfigure[SCHRISK]{\resizebox*{4.1cm}{!}{\includegraphics{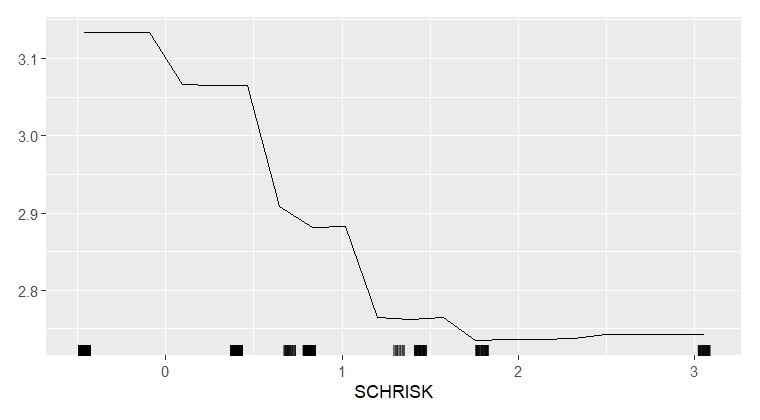}}}
    \hspace{0.3cm}
    \subfigure[BULLIED]{\resizebox*{4.1cm}{!}{\includegraphics{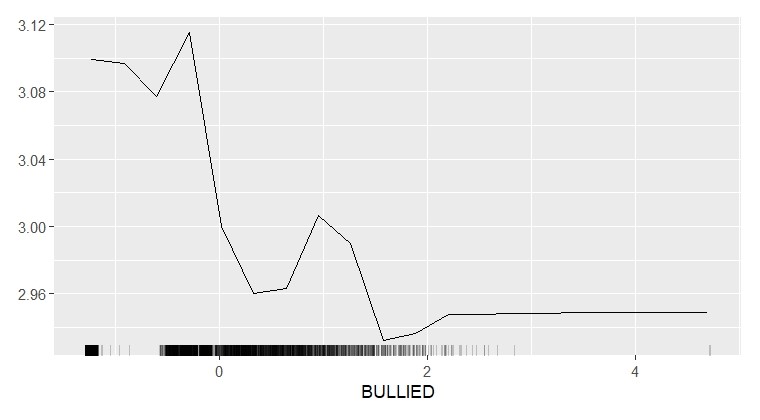}}}
    \hspace{0.3cm}
    \subfigure[study\_time]{\resizebox*{4.1cm}{!}{\includegraphics{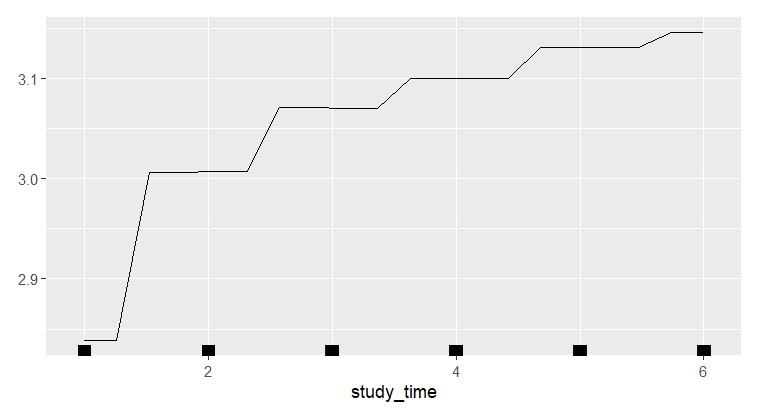}}}
    \hspace{0.3cm}
    \subfigure[TEACHSUP]{\resizebox*{4.1cm}{!}{\includegraphics{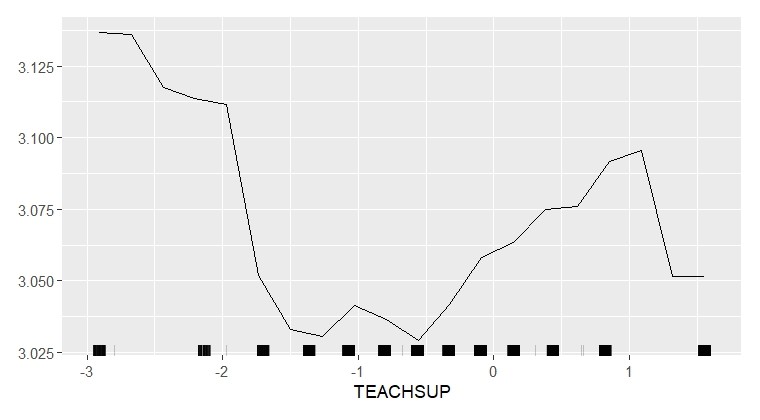}}}
    \hspace{0.3cm}
    \subfigure[video\_games]{\resizebox*{4.1cm}{!}{\includegraphics{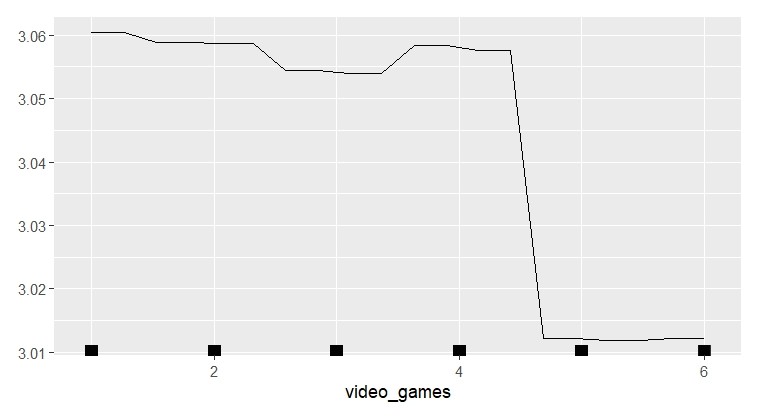}}}
    \hspace{0.3cm}
   \subfigure[grade]{\resizebox*{4.1cm}{!}{\includegraphics{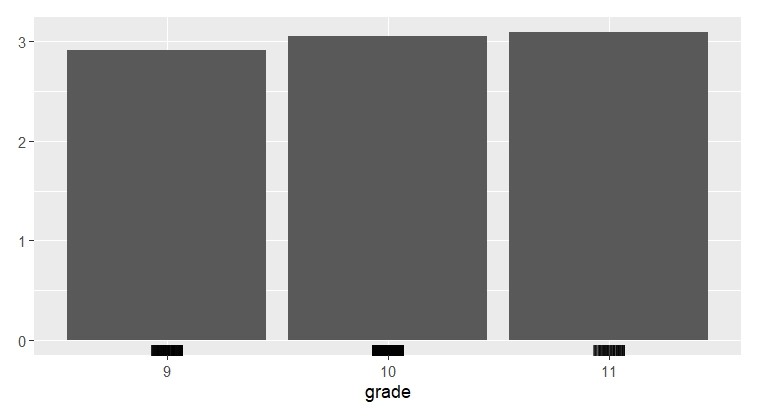}}}
    \hspace{0.3cm}
    \subfigure[internet\_availability]{\resizebox*{4.1cm}{!}{\includegraphics{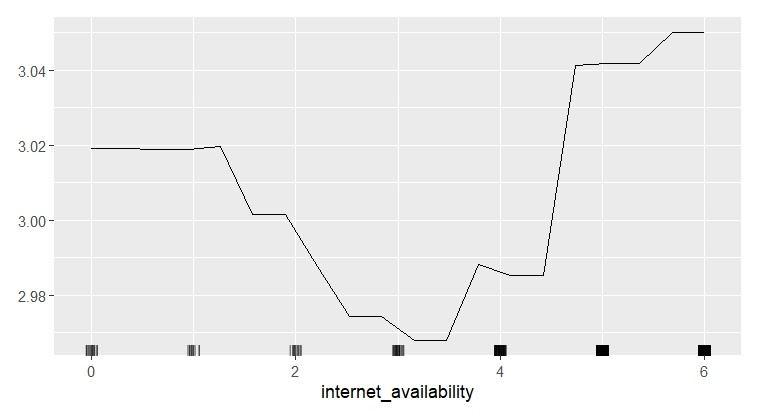}}}
    \hspace{0.3cm}
    \subfigure[immig]{\resizebox*{4.1cm}{!}{\includegraphics{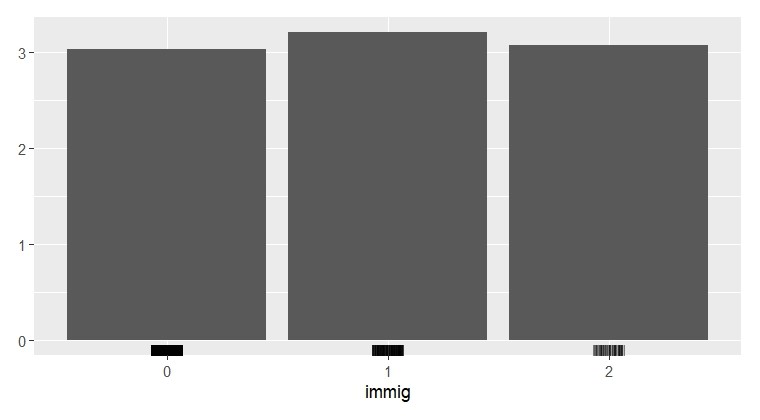}}}
    \caption{Partial Plots for the fixed component of OMERF, in the real-world case study. The \(y\)-axis reports the increment/decrement of the target variable of the random forest in the iterative procedure, given the covariate on the \(x\)-axis.} \label{fig:par_plot_case}
\end{figure}

%\begin{figure}[H]
%    \centering
%    \includegraphics[width=\textwidth]{Images/parplot_new.png}
%    \caption{Partial Plots for the fixed component of OMERF, in the real-world case study.  \CM{INGRANDIREI LA FIGURA A TUTTA PAGINA E METTEREI MAX 4 PP PER RIGA}\LR{Giulia sta sistemando l'asse y del plot RICORDIAMOCI DI SPECIFICARE COSA C'è SULL'ASSE Y}}
%    \label{fig:par_plot_case}
%\end{figure}

%\CM{Sarebbe interessante a questo punto paragonare altre covariate tra CLMM e OMERF per vedere se OMERF ci rivela qualcosa (non linear) che nel CLMM non si vede! così da affiancare ESCS a qualche altra var da commentare e raccontare un po' di più il caso studio}

%%ci interessa questo commento?
%{As the variance of the standard logistic distribution is \(\pi^2 / 3 \cong 3.29\), and} 
The estimated variance of the random effects results to be \(\hat{\sigma}^2 = 1.695\) for OMERF, leading to the following Intraclass Correlation Coefficients (ICC) \citep{grilli2011multilevel} for the underlying linear model:
\begin{gather*}
    \begin{aligned}
        ICC = \frac{\sigma^2}{\sigma^2+\pi^2 / 3}= 0.340.
    \end{aligned}
\end{gather*}
This value of ICC, measuring the unexplained variance in the response that can be attributed to the nested structure of students,
% \CM{FORSE TOGLIEREI QUESTO: together with the fact that some random intercepts are significantly different from zero (Figure \ref{fig:reff})}
implies the existence of a substantial heterogeneity in the achievements among various classes. This highlight once again the importance of considering the hierarchical structure of these data.

Figure \ref{fig:reff} reports the estimated random intercepts, that in this case are interpreted as school-effects, of OMERF and CLMM, that result to be coherent. Their distribution around 0, together with the high ICC, confirms that the likelihood of a student to be in different proficiency levels, net to the effect of his/her personal characteristics, is influenced by the attended school.

%\textcolor{red}{PER CHIARA: NON SO BENE COSA SI POSSA COMMENTARE A RIGUARDO: STESSA/SIMILE SHAPE? FORSE SAREBBE BELLO AVERE ANCHE UN VALORE NUMERICO DI ICC PER IL CLMM?}

%\CM{AGGIUNGEREI: CONFRONTO DELLE RANDOM INTERCPETS STIMATE TRA CLMME E OMERF E CONFRONTO DELLA PARTE FISSA PER ENFATIZZARE OMERF. LA PROVA TOGLIENDO ESCS A COSA PORTA? ERA STATA FATTA?}

\begin{figure}
\centering
    \subfigure[OMERF methos]{\resizebox*{7cm}{!}{\includegraphics{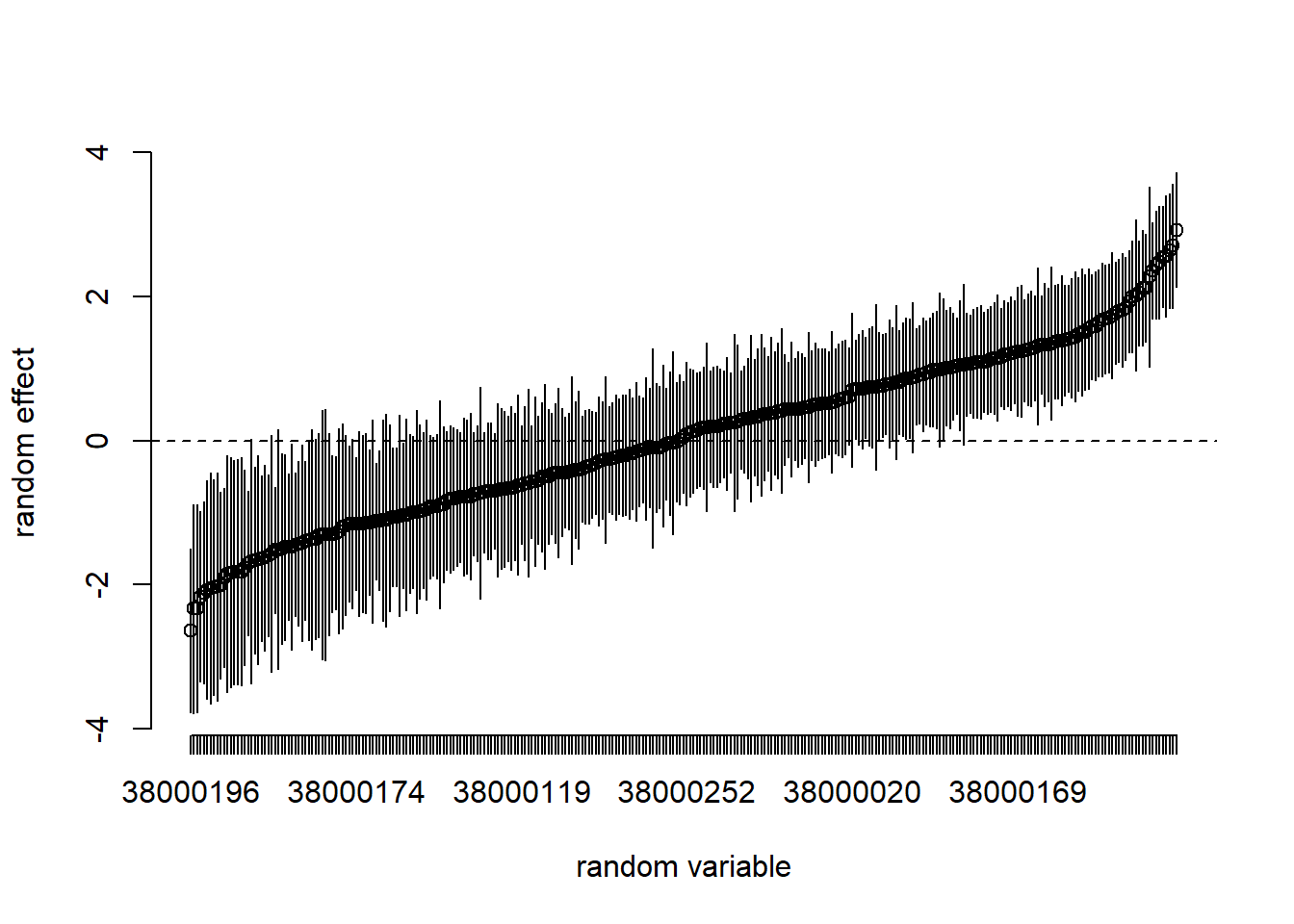}}}\hspace{0.5cm}
    \subfigure[CLMM method]{\resizebox*{7cm}{!}{\includegraphics{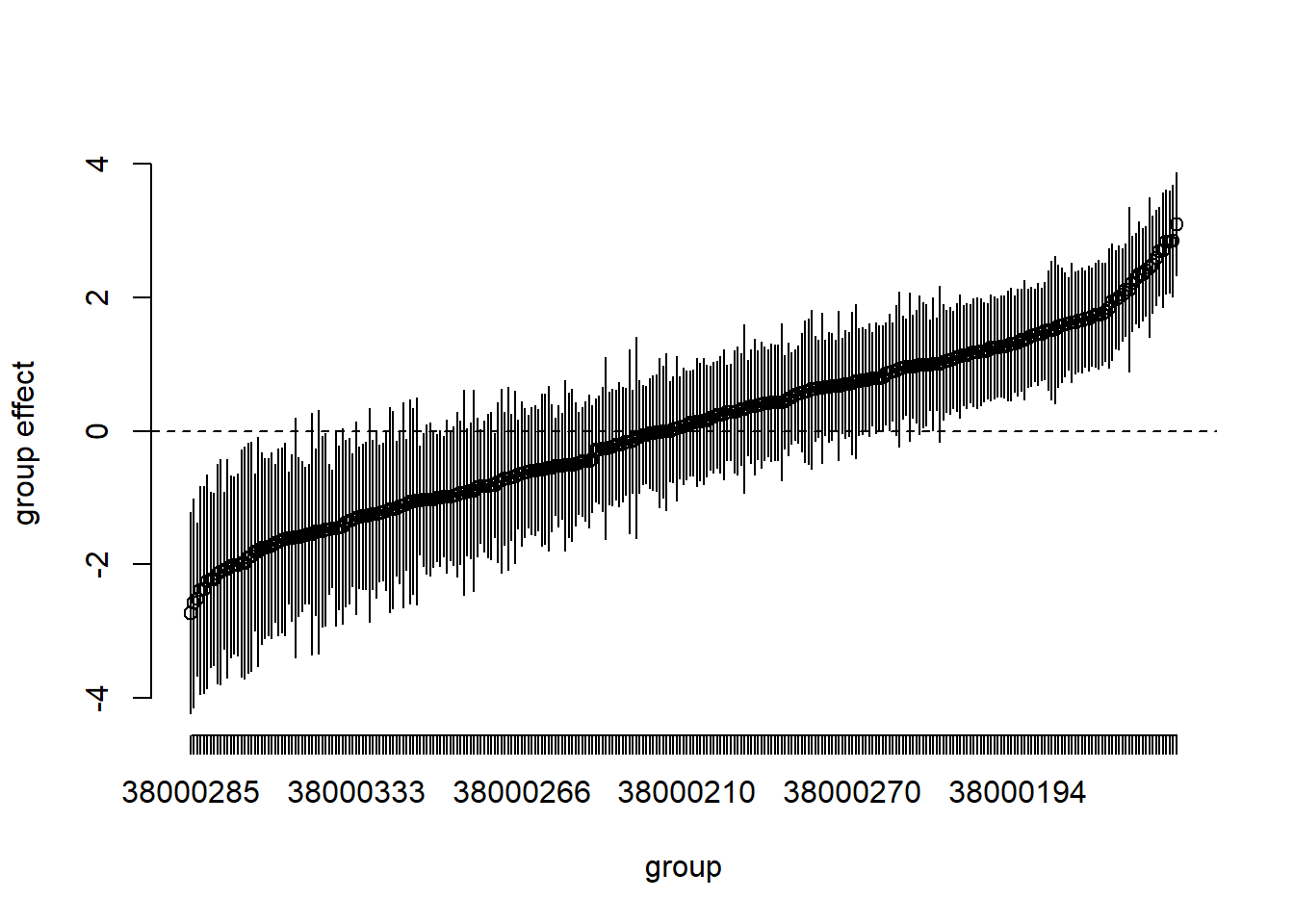}}}
    \caption{Estimated Random intercepts with their $95\%$ confidence intervals} \label{fig:reff}
\end{figure}

%-----------------------------------------------------------------------------
% CONCLUSION
%-----------------------------------------------------------------------------
\section{Conclusions}
\label{sec:conclusions}
\color{black}
In this study, we propose the Ordinal Mixed-Effects Random Forest (OMERF), a novel method that expands the utility of random forests to analyze hierarchical data with ordinal response variables. OMERF adopts the framework of cumulative linear mixed models but employs a random forest to estimate the fixed component. By doing so, it inherits the flexibility and predictive power of random forests while preserving the structure of mixed-effects models for ordinal response.
This novel approach makes a valuable contribution to two branches of statistical literature: the one on tree-based mixed-effects models and the one on ordinal models.

%{To show the performance of the proposed model simulation study is performed. Subsequently, OMERF algorithm is applied to a real world dataset to predict mathematics performance of Italian students in PISA 2022 test.} %containing  all the personal information stored in the school's electronic register of a prestigious high school in Milan, in order to identify students who may be at risk of failure and generally of predicting their academic progress.
A simulation study shows that OMERF outperforms the counterpart methods when the fixed component 
exhibits complex non linear structures, in presence of both light and strong heterogeneity at the group level, while linear models still perform better in case of linear predictors. 

%In the simulation study, the performance of the proposed mixed-effects tree-based method was a clear improvement compared to CLM and CLMM models when the data generating process features nonlinear patterns. In addition, the predictive power of the OMERF model was comparable to the one of the ordinal random forest model.

When applied to a real-world case study to predict mathematics proficiency levels of Italian students in the PISA 2022 test, OMERF performs similarly
but slightly worse than CLMM, in terms of predictive power. Indeed, the strong linear relationship of the most important predictor (ESCS) with the response favors the linear model. Nonetheless, OMERF offers the advantage of investigating potential interactions and non linearities of other covariates, % \CM{DA AGGIUSTARE DOPO IL CASE STUDY}
still disentangling the school effect on student outcomes, that, in the educational data mining context, as in many other contexts, is essential to properly understand students and the settings in which they learn.  

%{Overall, the main advantages of the OMERF algorithm are its flexibility and data inspection capability.
%By relaxing the linear assumption of the fixed-effects part, the method could model more complex functional forms, detecting also potential interactions
%among covariates, for both categorical and continuous variables.
%For this reason, the decision to use OMERF over CLMM depends on the complexity of the fixed effects structure, which is often not known in advance.
%If fixed-effect covariates exhibit a linear association with the response and are not correlated, CLMM is expected to perform better.
%Conversely, if the covariate set is large, and the covariates potentially interact, creating complex patterns in their association with the response, OMERF is expected to outperform parametric methods with predefined functional forms.
%At the same time, when data present a hierarchy, the method is able to take into account the dependence structure within observations and to model it.
%In the educational data mining context, this aspect is essential in order to better understand students and the settings in which they learn.}

In conclusion, OMERF proves to be a powerful and an easily interpretable method, that can deal with grouped data structures and can be applied to face complex real data challenges. Future developments might consider to enrich the modeling of the fixed-effect part by incorporating both a linear component and a tree-based component (see, for example, \cite{gottard2019tree}). Moreover, OMERF should be employed to deal with more complex real-world data to test its efficacy in absence of clear linear predictors.  

%{In the case study, we give a contribution to the learning analytics area. OMERF method, when applied to educational data, can be a useful tool to support the definition of best practices and
%new tutoring programmes aimed at enhancing student performances and helping students at risk. }

%\CM{Aggiungere la questione di incorporare sia linear che RF nei fixed, parlare di limitations and future }

%However, a potential limitation of the OMERF algorithm stems from the initial step where, during the first initialization, the method estimates the target values (\(\eta_{ijc}\)) through a CLM.
%his model imposes a linear effect of covariates on a transformation of the response thereby potentially losing the intricacies that the random forest could capture in the subsequent step of the algorithm.
%Therefore, a hint for further developments might be to consider using a different model that can successfully capture nonlinear relationships and interactions between covariates and help initialize the \(\eta_{ijc}\) estimates, thereby allowing the OMERF method to perform at its best.

\vspace{1cm}
\textbf{Acknowledgements}: The present research is part of the activities of \lq\lq Dipartimento di Eccellenza 2023-2027".
$\,$
\bibliographystyle{apalike}
\bibliography{bibliography.bib}

\begin{thebibliography}{}

\bibitem[Agresti, 2010]{agresti2010analysis}
Agresti, A. (2010).
\newblock {\em Analysis of ordinal categorical data}, volume 656.
\newblock John Wiley \& Sons.

\bibitem[Ananth and Kleinbaum, 1997]{ananth1997regression}
Ananth, C.~V. and Kleinbaum, D.~G. (1997).
\newblock Regression models for ordinal responses: a review of methods and applications.
\newblock {\em International journal of epidemiology}, 26(6):1323--1333.

\bibitem[Ballante et~al., 2022]{ballante2022new}
Ballante, E., Figini, S., and Uberti, P. (2022).
\newblock A new approach in model selection for ordinal target variables.
\newblock {\em Computational Statistics}, 37(1):43--56.

\bibitem[Breiman, 2001]{breiman2001random}
Breiman, L. (2001).
\newblock Random forests.
\newblock {\em Machine learning}, 45:5--32.

\bibitem[Breiman, 2017]{breiman2017classification}
Breiman, L. (2017).
\newblock {\em Classification and regression trees}.
\newblock Routledge.

\bibitem[Cardoso and Sousa, 2011]{cardoso2011measuring}
Cardoso, J.~S. and Sousa, R. (2011).
\newblock Measuring the performance of ordinal classification.
\newblock {\em International Journal of Pattern Recognition and Artificial Intelligence}, 25(08):1173--1195.

\bibitem[Christensen, 2022]{ordinal}
Christensen, R. H.~B. (2022).
\newblock ordinal---regression models for ordinal data.
\newblock R package version 2022.11-16. https://CRAN.R-project.org/package=ordinal.

\bibitem[Cohen, 1960]{cohen1960coefficient}
Cohen, J. (1960).
\newblock A coefficient of agreement for nominal scales.
\newblock {\em Educational and psychological measurement}, 20(1):37--46.

\bibitem[de~Raadt et~al., 2021]{de2021comparison}
de~Raadt, A., Warrens, M.~J., Bosker, R.~J., and Kiers, H.~A. (2021).
\newblock A comparison of reliability coefficients for ordinal rating scales.
\newblock {\em Journal of Classification}, pages 1--25.

\bibitem[Fokkema et~al., 2018]{fokkema2018detecting}
Fokkema, M., Smits, N., Zeileis, A., Hothorn, T., and Kelderman, H. (2018).
\newblock Detecting treatment-subgroup interactions in clustered data with generalized linear mixed-effects model trees.
\newblock {\em Behavior research methods}, 50:2016--2034.

\bibitem[Fontana et~al., 2021]{fontana2021performing}
Fontana, L., Masci, C., Ieva, F., and Paganoni, A.~M. (2021).
\newblock Performing learning analytics via generalised mixed-effects trees.
\newblock {\em Data}, 6(7):74.

\bibitem[Gaudette and Japkowicz, 2009]{gaudette2009evaluation}
Gaudette, L. and Japkowicz, N. (2009).
\newblock Evaluation methods for ordinal classification.
\newblock In {\em Advances in Artificial Intelligence: 22nd Canadian Conference on Artificial Intelligence, Canadian AI 2009 Kelowna, Canada, May 25-27, 2009 Proceedings 22}, pages 207--210. Springer.

\bibitem[Goldfeld and Wujciak-Jens, 2020]{simstudy}
Goldfeld, K. and Wujciak-Jens, J. (2020).
\newblock simstudy: Illuminating research methods through data generation.
\newblock {\em Journal of Open Source Software}, 5(54):2763.

\bibitem[Gottard et~al., 2019]{gottard2019tree}
Gottard, A., Grilli, L., Rampichini, C., Vannucci, G., et~al. (2019).
\newblock Tree embedded linear mixed models.
\newblock {\em Book of Short Papers}, page 239.

\bibitem[Grilli and Rampichini, 2011]{grilli2011multilevel}
Grilli, L. and Rampichini, C. (2011).
\newblock Multilevel models for ordinal data.
\newblock {\em Modern analysis of customer surveys: With applications using R}, pages 391--411.

\bibitem[Hajjem et~al., 2011]{hajjem2011mixed}
Hajjem, A., Bellavance, F., and Larocque, D. (2011).
\newblock Mixed effects regression trees for clustered data.
\newblock {\em Statistics \& probability letters}, 81(4):451--459.

\bibitem[Hajjem et~al., 2014]{hajjem2014mixed}
Hajjem, A., Bellavance, F., and Larocque, D. (2014).
\newblock Mixed-effects random forest for clustered data.
\newblock {\em Journal of Statistical Computation and Simulation}, 84(6):1313--1328.

\bibitem[Hajjem et~al., 2017]{hajjem2017generalized}
Hajjem, A., Larocque, D., and Bellavance, F. (2017).
\newblock Generalized mixed effects regression trees.
\newblock {\em Statistics \& Probability Letters}, 126:114--118.

\bibitem[Hornung, 2020]{hornung2020ordinal}
Hornung, R. (2020).
\newblock Ordinal forests.
\newblock {\em Journal of Classification}, 37:4--17.

\bibitem[Hubert and Arabie, 1985]{hubert1985comparing}
Hubert, L. and Arabie, P. (1985).
\newblock Comparing partitions.
\newblock {\em Journal of classification}, 2:193--218.

\bibitem[James et~al., 2013]{james2013introduction}
James, G., Witten, D., Hastie, T., Tibshirani, R., et~al. (2013).
\newblock {\em An introduction to statistical learning}, volume 112.
\newblock Springer.

\bibitem[Liaw and Wiener, 2002]{RF}
Liaw, A. and Wiener, M. (2002).
\newblock Classification and regression by randomforest.
\newblock {\em R News}, 2(3):18--22.

\bibitem[McCullagh, 1980]{mccullagh1980regression}
McCullagh, P. (1980).
\newblock Regression models for ordinal data.
\newblock {\em Journal of the Royal Statistical Society: Series B (Methodological)}, 42(2):109--127.

\bibitem[Pellagatti et~al., 2021]{pellagatti2021generalized}
Pellagatti, M., Masci, C., Ieva, F., and Paganoni, A.~M. (2021).
\newblock Generalized mixed-effects random forest: A flexible approach to predict university student dropout.
\newblock {\em Statistical Analysis and Data Mining: The ASA Data Science Journal}, 14(3):241--257.

\bibitem[Pinheiro and Bates, 2006]{pinheiro2006mixed}
Pinheiro, J. and Bates, D. (2006).
\newblock {\em Mixed-effects models in S and S-PLUS}.
\newblock Springer science \& business media.

\bibitem[{R Core Team}, 2022]{rlanguage}
{R Core Team} (2022).
\newblock {\em R: A Language and Environment for Statistical Computing}.
\newblock R Foundation for Statistical Computing, Vienna, Austria.

\bibitem[Raudenbush and Willms, 1995]{raudenbush1995estimation}
Raudenbush, S.~W. and Willms, J. (1995).
\newblock The estimation of school effects.
\newblock {\em Journal of educational and behavioral statistics}, 20(4):307--335.

\bibitem[Sela and Simonoff, 2012]{sela2012re}
Sela, R.~J. and Simonoff, J.~S. (2012).
\newblock Re-em trees: a data mining approach for longitudinal and clustered data.
\newblock {\em Machine learning}, 86:169--207.

\bibitem[Shashua and Levin, 2002]{shashua2002ranking}
Shashua, A. and Levin, A. (2002).
\newblock Ranking with large margin principle: Two approaches.
\newblock {\em Advances in neural information processing systems}, 15.

\bibitem[Speiser et~al., 2020]{speiser2020bimm}
Speiser, J.~L., Wolf, B.~J., Chung, D., Karvellas, C.~J., Koch, D.~G., and Durkalski, V.~L. (2020).
\newblock Bimm tree: a decision tree method for modeling clustered and longitudinal binary outcomes.
\newblock {\em Communications in Statistics-Simulation and Computation}, 49(4):1004--1023.

\bibitem[Tutz, 2003]{tutz2003generalized}
Tutz, G. (2003).
\newblock Generalized semiparametrically structured ordinal models.
\newblock {\em Biometrics}, 59(2):263--273.

\bibitem[Tutz, 2022]{tutz2022ordinal}
Tutz, G. (2022).
\newblock Ordinal trees and random forests: Score-free recursive partitioning and improved ensembles.
\newblock {\em Journal of Classification}, 39(2):241--263.

\bibitem[Tutz and Hennevogl, 1996]{tutz1996random}
Tutz, G. and Hennevogl, W. (1996).
\newblock Random effects in ordinal regression models.
\newblock {\em Computational Statistics \& Data Analysis}, 22(5):537--557.

\bibitem[Yang et~al., 2020]{yang2020algorithm}
Yang, Y., Chen, B., and Yang, Z. (2020).
\newblock An algorithm for ordinal classification based on pairwise comparison.
\newblock {\em Journal of Classification}, 37:158--179.

\end{thebibliography}

% \begin{thebibliography}{99}

% \bibitem{1} Spiegel, M. R. (1981). Theory and problems of Advanced Calculus: Si (metric) edition. McGraw-Hill. 

% \end{thebibliography}

\end{document}